\documentclass[useAMS,usenatbib,fleqn]{mn2e}
\usepackage{graphicx}
\usepackage{epstopdf}
\usepackage[figuresright]{rotating}
\usepackage{subfigure,color}
\usepackage{longtable}
\usepackage{tabularx,multicol,lipsum}
\usepackage{xspace}

\usepackage{hyperref}%
\usepackage[T1]{fontenc}
\usepackage{ae,aecompl}
\usepackage{longtable}
\usepackage{placeins}
\usepackage{multicol}
\usepackage{subfigure}
\usepackage{siunitx}

\newcommand{\ion}[2]{{\textrm{#1}}{\textrm{\sc #2}}}

\definecolor{pink}{rgb}{.9,.2,.5}  
\definecolor{purple}{rgb}{.5,.6,.7}

\def\la{\raise.5ex\hbox{$<$}\kern-.8em\lower 1mm\hbox{$\sim$}}
\def\ma{\raise.5ex\hbox{$>$}\kern-.8em\lower 1mm\hbox{$\sim$}}

\def\kms{$\rm km\, s^{-1}$}
\def\cm3{$\rm cm^{-3}$}

\def\Vs{$V_{\rm s}$}
\def\n0{$n_{0}$}
\def\B0{$B_{0}$}

\def\Te{$T_{\rm e}$}

\def\L12{L$_{12\mu m}$}
\def\F12{F$_{12\mu m}$}

\def\Hb{H${\beta}$}

\def\Hg{H$\gamma$}

\title[Composite models  for Seyfert~2 AGNs]{Chemical abundances of  Seyfert 2 AGNs-- IV.
 Composite models  calculated by photoionization + shocks
 }

\author[Dors et al.]
            {O.~L. Dors$^{1}$\thanks{E-mail:olidors@univap.br}, M. Contini$^{2}$, R.~A. Riffel$^{3}$, E.~ P\'erez-Montero$^{4}$, A.~C. Krabbe$^{1}$,  
	     \newauthor{M.~V.~Cardaci$^{5,6}$, G.~F.~H\"{a}gele$^{5,6}$}\\
$^{1}$ Universidade do Vale do Para\'iba, Av. Shishima Hifumi, 2911, Cep
12244-000, S\~ao Jos\'e dos Campos, SP, Brazil\\ 
$^{2}$ School of Physics and Astronomy, Tel Aviv University, Tel Aviv 69978, Israel \\
$^{3}$ Universidade Federal de Santa Maria, Av. Roraima 1000, Cep 97105-900, Santa Maria, Brazil \\
$^{4}$ Instituto de Astrof{\'i}sica de Andaluc{\'i}a, Camino Bajo de Hu{\'e}tor s/n, Aptdo. 3004, E18080-Granada, Spain. \\
$^{5}$ Instituto de Astrof\'{i}sica de La Plata (CONICET-UNLP), Argentina \\
$^{6}$ Facultad de Ciencias Astron\'{o}micas y Geof\'{i}sicas, Universidad Nacional de La Plata, Paseo del Bosque s/n, 1900 La Plata, Argentina \\
} 
\begin{document}

\date{Accepted 2020 Month  00. Received 2020 Month 00; in original form 2019 December 17}

\pagerange{\pageref{firstpage}--\pageref{lastpage}} \pubyear{2011}

\maketitle

\label{firstpage}

\begin{abstract}

We build detailed composite models of photoionization and shock ionization based on
the {\sc SUMA} code to reproduce emission lines  emitted  from the
Narrow Line Regions (NLR) of Seyfert~2 nuclei. The aim of this work is to investigate diagram AGN positions according to shock parameters,
shock effects on  the gas temperature and ionization structures and
derive  a semi-empirical abundance calibration based on emission-line ratios  little sensitive to the shock presence. 
The models were used to reproduce 
optical ($3000 \: < \: \lambda$(\AA) $< \: 7000$) emission line intensities of 244 local ($z \: \la \: 0.4$)
Seyfert 2s, whose observational data were selected from  Sloan Digital Sky Survey DR7.
Our models suggest  that  shocks in  Seyfert~2 nuclei  have velocities in the
range of  50-300 $\rm km \: s^{-1}$ and  imply a narrower metallicity range ($0.6 \: \la \:  (Z/Z_{\odot}) \: \la \: 1.6$)
than those derived using pure photoionization models. Our results  indicate that 
shock velocity in AGNs can not be estimated using standard optical line ratio diagrams, based on integrated spectra.
Our  models predict a  different temperature structure and $\rm O^{+}$/O and $\rm O^{2+}$/O fractional abundances  throughout the  NLR clouds
than those derived from  pure photoionization models,  mainly in shock-dominated objects.
This suggests that, in order to minimize the shock effects, 
the combination of emission-lines emitted by ions with similar intermediate  ionization potential could be good metallicity indicators. 
Finally, we derive two calibrations between the N/O abundance ratio
and the $N2O2$=log([\ion{N}{ii}]$\lambda$6584/[\ion{O}{ii}]$\lambda$3727) and $N2$=log([\ion{N}{ii}]$\lambda$6584/H$\alpha$) indexes 
which agree with that
derived from pure photoionization models.
\end{abstract}
\begin{keywords}
galaxies: active  -- galaxies:  abundances -- galaxies: evolution -- galaxies: nuclei --
galaxies: formation-- galaxies: ISM -- galaxies: Seyfert
\end{keywords}


\section{Introduction}
\label{intro}

 Active Galactic Nuclei (AGNs) and Star-Forming Regions (SFs) present in their
spectra prominent emission lines observed from  X-ray  to radio wavelengths.
The relative intensities  and the  profiles of these lines reveal the
 properties of the gas phase, such as  chemical
abundances, ionization degree, kinematics, etc. 
Since AGNs and SFs are thought to be ubiquitous in the
Universe from their very first stages, investigating the
physics underlying these objects is crucial to understand their role in the cosmic evolution of galaxies.

The seminal paper by \citet{baldwin81}  allowed the first 
taxonomy of emitter objects through diagnostic diagrams (hereafter  BPT diagrams) containing
optical emission-line ratios (see also \citealt{veilleux87, kewley01, kauffmann03, enrique13}). 
BPT diagrams show that AGNs (and most planetary nebulae, see \citealt{frew10})  exhibit 
higher line ratios (e.g. [\ion{O}{iii}]$\lambda$5007/H$\beta$ and [\ion{N}{ii}]$\lambda$6584/H$\alpha$)
than those of SFs. The difference between line intensity ratios of AGNs, on the scenario
of  photoionization due to radiation emitted by gas accretion into a supermassive black hole (SMBH), 
is mainly due to a much harder ionizing
spectral energy distribution (SED) in combination
with a higher ionization degree (e.g. \citealt{enrique19}) 
and to a larger metallicity in AGN hosts (e.g. \citealt{grazyna84, thaisa90, thaisa98, groves06, feltre16}). 
However, high values of line intensity ratios are also obtained   adopting shock dominated models with relatively high shock 
velocities (\Vs), because the higher \Vs\, the higher  [\ion{O}{iii}]$\lambda$5007/H$\beta$  ratio 
(e.g. \citealt{spence16}). Therefore, in a more realistic physical
frame, models built to reproduce observational line intensity ratios
should  account for a composite  ionization source (AGN+shock),  which lead to   
more reliable results than those obtained by pure photoionization or pure shock models.

Collisional ionization and heating of  the  gas by  the shocks  contribute to
the  line intensities measured in  the  spectra of both AGN hosts and SF galaxies \citep{aldrovandi84, dopita95, dopita96}.
Shocks have a strong influence on the gas properties
(e.g. \citealt{viegas89, dopita96, allen08}), in particular, on 
 the gas density, temperature downstream,  
  cooling rates throughout the clouds, etc. Consequently, they affect the
 emission lines. The element abundances,  the flux from the active centre (AC),  the dust-to-gas ratios, etc,  characterise the gas in  pre-shock regions.
Shock velocities  (or a more turbulent gas) generated
by outflows in AGNs (e.g. \citealt{rosario10, riffel14,wylezalek20}) are  higher than those in SF galaxies which originate mainly from
stellar winds of  young massive stars (e.g. \citealt{dyson79, westmoquette07, rozas07, amorin12, bosch19}).
Therefore, the optical line profiles observed, for instance, in narrow line regions (NLRs) of Seyfert 2 AGNs show
Full Width at Half Maximum (FWHM) 
ranging from 200 to 1000 \kms (e.g. \citealt{koski78,  vaona12, zhang13}) while those  measured in 
SFs regions are  $\la \: 200 \rm \: km \: s^{-1}$ (e.g. \citealt{melnick77, skillmann84, relano05, hagele13, bresolin20}).
The FWHM of optical emission lines is a good tracer of the shock velocities (e.g. \citealt{contini12}).

Regarding the metallicity, shocks in NLRs can be one of the causes 
behind the $T_{\rm e}$-problem in Seyfert~2 nuclei. In fact, \citet{dors15} showed that the determination  of the
metallicity $Z$ (in terms of the O/H abundance)  from  the direct measurement of the 
electron temperature\footnote{This method is referred
as  $T_{\rm e}$-method or direct method.} 
 ($T_{\rm e}$),  provides a reliable  -- even if approximated --  method for SFs (see, e.g. \citealt{pilyugin03, hagele06, hagele08, contini14}) but 
  produces unrealistic low  $Z$
(see also \citealt{dors20a}) in the NLRs of Seyfert 2 galaxies. 
Sub-solar {\rm metallicities} are obtained  as a consequence of the high  values of the electron temperature ($T_{\rm e} \: > \: 20\,000$ K) 
in the NLRs which translates into  low $Z$. \citet{heckman79} pointed out that 
such high temperatures require another  source of energy in addition 
to photoionization,   e.g. the presence of  shocks (see also \citealt{zhang13, contini17}). 
\citet{dors20b} presented a new formalism of the $T_{\rm e}$-method for Seyfert~2, i.e.
a new relation between the temperature of the low ($t_{2}$)  and high  ($t_{3}$) ionization gas zones,
 which is  different from  those commonly used for chemical abundance studies in the \ion{H}{ii} regions  
(e.g. \citealt{garnett92, hagele08, enrique14}). 
Despite the use of this new methodology produces a cut down of the difference between the O/H abundances es-
timated through the $T_{\rm e}$-method and those obtained by pure
photoionization models to  $\sim 0.2$ dex, some caveats still prevent
the use of the $T_{\rm e}$-method for AGNs.  For instance, \Te\, can be calculated from
$R_{\rm O3}$=([\ion{O}{iii}]$(\lambda 4959+ \lambda 5007)/\lambda4363$) line ratio
in the range of $700 \: \ga \: R_{\rm O3}  \: \ga \: 30$ 
which corresponds to  $7000 \: \la \: T_{\rm e} (\rm K) \: \la \: 23\:000$ \citep{hagele08}. 
However, $R_{\rm O3}$  lower than 30 
is derived in some Seyfert~2 (e.g. \citealt{komossa97, nagao01, vaona12}), 
indicating  $T_{\rm e} \: > \: 23\,000$ K and a 
limited use of the $T_{\rm e}$-method for this class of objects. 
Moreover, the discrepancy between  $T_{\rm e}$  calculated from  measurements
of $R_{\rm O3}$ and those predicted by pure photoionization models 
is systematic,  meaning that it increases  (from $\sim 0$ to 
$\sim  11\:000$ K) when the values derived by the $T_{\rm e}$-method increase  
(varying in the 11\,000--20\,000 range; \citealt{dors20b}).
This indicates that  another mechanism is acting in NLRs.
  
Shocks with velocities  $V_s \: \la \: 400 \rm \: km \: s^{-1}$ have been proposed to be at work  in the NLR
of Seyfert~2s galaxies (e.g. \citealt{contini17}). 
High \Te\, throughout the   emitting clouds are mainly due, as previously reported,  to the presence of shocks
(e.g. \citealt{dors15, contini17}).
They can produce some uncertainties in the use of the $T_{\rm e}$-method.
A basic difference between radiation dominated  and shock dominated models consists in the profile of \Te\,
in the recombination region of the gas within the emitting nebula because the cooling rate downstream of the shock front
is strengthened by compression \citep{contini17}.   

The main goal of this paper is to investigate shock effects on the NLR gas of Seyfert~2 galaxies,  
 analysing the  loci of the corresponding line ratios  in BPT  diagrams according to shock parameters and abundances,
 investigating  the influence of shocks on temperature and ionization structure
and deriving an abundance calibration based on emission-line ratios less sensitive  to shock.
 Therefore, we adopt composite  models (photoionization + shock) using the {\sc SUMA} code \citep{viegas89} 
in order to reproduce the  optical narrow emission-lines of  244 Seyfert~2 nuclei whose data were
taken from the Sloan Digital Sky Survey (SDSS, \citealt{york00}) by \citet{dors20a}.
This paper is organized as follows: in Section~\ref{meth}
the methodology (observational data and models)  is presented; 
in Section~\ref{res} the results of  detailed modelling of the spectra are presented, while 
the discussion and conclusion
remarks are given in Sections~\ref{disc}  and \ref{conc}, respectively.

  
\section{Methodology}
\label{meth}
\subsection{Observational data}
\label{obs}

We consider optical narrow emission line intensities ($3600 \: < \: \lambda$(\AA) $< \: 7200$) 
of a sample  of Seyfert~2  galaxies compiled by \citet{dors20a}. 
These data were taken from SDSS-DR7 \citep{york00, abazajian09}  
and the emission line intensity measurements are made available by the MPA/JHU group\footnote{https://wwwmpa.mpa-garching.mpg.de/SDSS/DR7/}.
To select the objects, \citet{dors20a} applied the criteria to  separate AGN-like and SF-like objects proposed
by \citet{kewley01, kewley06} and \citet{enrique13} and based on BPT diagrams. After selecting a sample of AGNs, 
\citet{dors20a} carried out a cross-correlation between  basic information from the
 SDSS-DR7 and in  NED/IPAC\footnote{ned.ipac.caltech.edu}
(NASA/IPAC Extragalactic Database) catalogues
in order to  obtain only Seyfert~2 AGNs. This procedure eliminates from
the sample SFs, star-forming galaxies, Seyfert 1 galaxies, quasars
and Planetary Nebulae.

The resulting sample consists of 463 Seyfert~2  AGNs with redshifts $z \: \la \: 0.4$ and with stellar 
masses of the host galaxies
(also taken from the MPA-JHU group) in the range of $9.4 \: \la \: \log(M/ \rm M_{\odot}) \: \la \: 11.6$.
For our analysis we considered several emission-line intensities measured by the MPA-JHU group, reddening corrected 
and expressed in relation to H$\beta$, including 
[\ion{O}{ii}]$\lambda$3726+$\lambda$3729 (hereafter indicated as [\ion{O}{ii}]$\lambda$3727), 
[\ion{Ne}{iii}]$\lambda$3869, 
[\ion{O}{iii}]$\lambda$4363, 
[\ion{O}{iii}]$\lambda$5007, 
He\,I$\lambda$5876, 
[\ion{O}{i}]$\lambda$6300, 
H$\alpha$, 
[\ion{N}{ii}]$\lambda$6584, 
[\ion{S}{ii}]$\lambda$6716, 
[\ion{S}{ii}]$\lambda$6731, 
and [\ion{Ar}{iii}]$\lambda$7135 emission-lines. 
The reader is referred to \citet{dors20a} for a complete 
description of this sample.

\subsection{Composite models}
\label{model}
 
We built  models by using the  {\sc SUMA} code \citep{viegas89}   
in order to reproduce the observed  spectra of each object  of our sample. This code has the advantage of considering   a combination of two
ionization sources:  the photoionization flux from the AC and collisional effects from the shock. 
A detailed description of the input parameters is given by \citet{contini19} and 
a summary is presented in the following  paragraphs.

\begin{enumerate}
\item Ionization sources: The  composite effect of photoionization from the AC and collisional ionization and heating 
from  shocks are considered.
The radiation emitted by the AC is represented by  a power-law 
flux $F$  in number of photons cm$^{-2}$ s$^{-1}$ eV$^{-1}$ at the Lyman limit,
with  spectral indices  $\alpha_{\rm UV}= -1.5$ and $\alpha_X = -0.7$.
 The flux $F$ is measured at the innermost surface of the cloud, i.e. illuminated face of the
cloud. The shock input parameters are:   the shock velocity \Vs, the atomic pre-shock density \n0\ and
the  pre-shock magnetic field \B0, which defines the hydrodynamical field. They  are  used in the solution
of the Rankine-Hugoniot equations  at the shock front and downstream. These equations  are combined into the
compression equation which  leads to the  calculations of the density profile downstream.
We adopted  for all the models \B0=10$^{-4}$ G, which is suitable to the NLR of AGNs according to \citet{beck12}.
 It is worth mentioning  that the magnetic field \B0 has an important
role in models accounting for the shock. The stronger  \B0, the
lower the compression downstream. Therefore, lower densities are compensated
by a lower \B0.
The gas reaches a maximum
temperature in the immediate post-shock region $T_{\rm e}\sim$1.5$\times 10^5$ (\Vs/100 \kms)$^2$. 
 $T_{\rm e}$ decreases downstream following the cooling  rate of the gas.

\item Geometry: The models  adopt gaseous clouds in  a plane parallel geometry. The  geometrical thickness $D$ of the 
clouds  determines whether each model is radiation-bounded or matter-bounded.  $D$ 
is calculated consistently with the physical conditions and element abundances of the emitting gas.
\item Elemental abundances: The code considers, initially,  solar abundances \citep{grevesse88} for all the 
elements (H, He, C, N, O, Ne, Mg, Si, S, Cl, Ar and Fe), which are varied in the models  
 in order to reproduce the observed line ratios. 
We adopted the solar He/H relative abundance (in number of atoms, \citealt{ferland17}) of  0.1 for all the models. 
\item Dust: Dust  is present in the emitting clouds.  It is
 characterized by the dust-to-gas ratio $d/g$ and by the initial grain
radius $a_{\rm gr}$, which are  constrained by fitting the continuum spectral energy distribution (SED). 
We adopt in the modelling of the present AGN survey $d/g$ = 10$^{-14}$ by number and $a_{\rm gr}=1 \mu m$  
because   values in these ranges  lead to the best fit  of the AGN  SED
observed  through a wide range of wavelengths (see  e.g. \citealt{contini18}).
\end{enumerate}

\begin{figure*}
\centering
\includegraphics[angle=0,width=1.0\columnwidth]{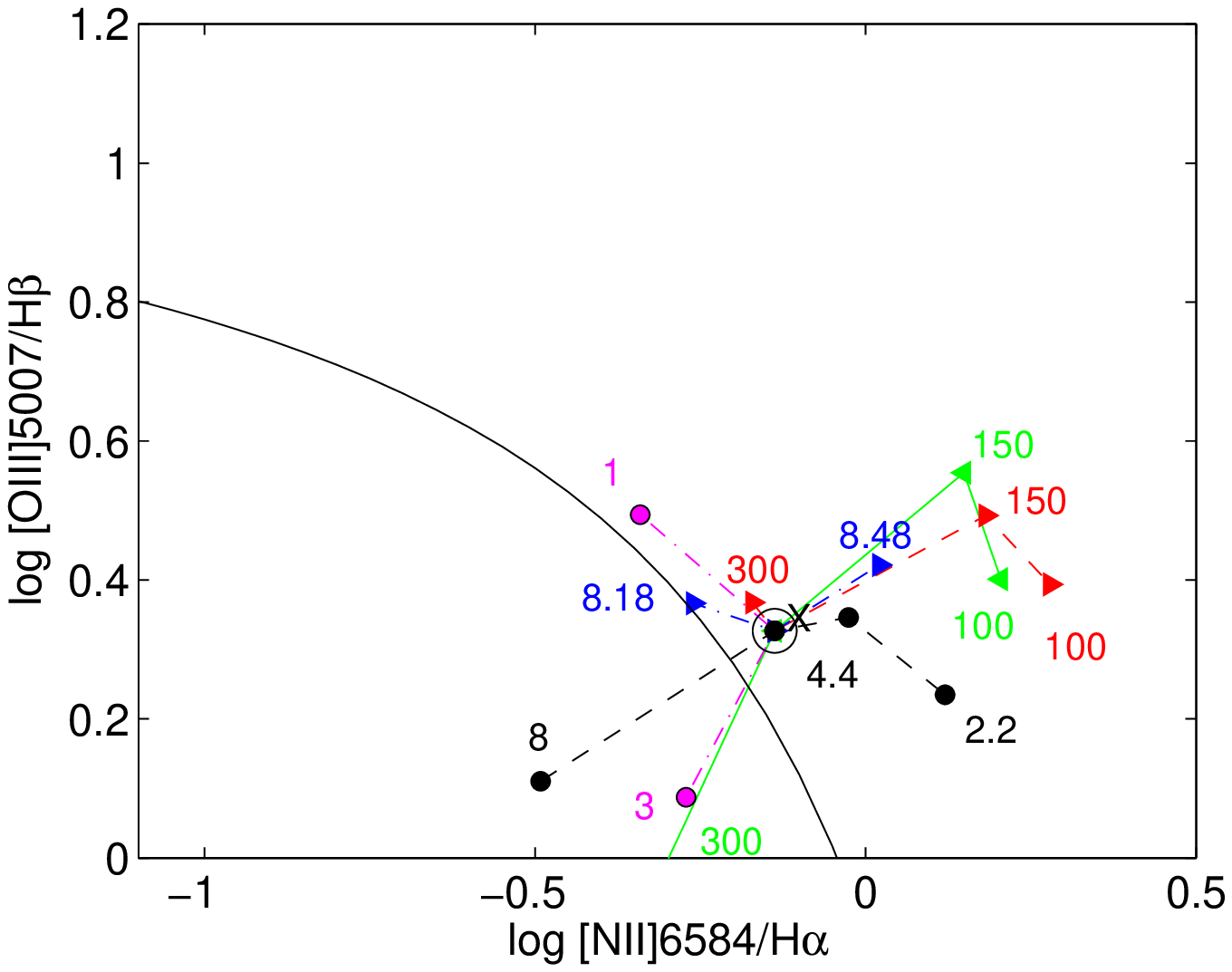} 
\includegraphics[angle=0,width=1.0\columnwidth]{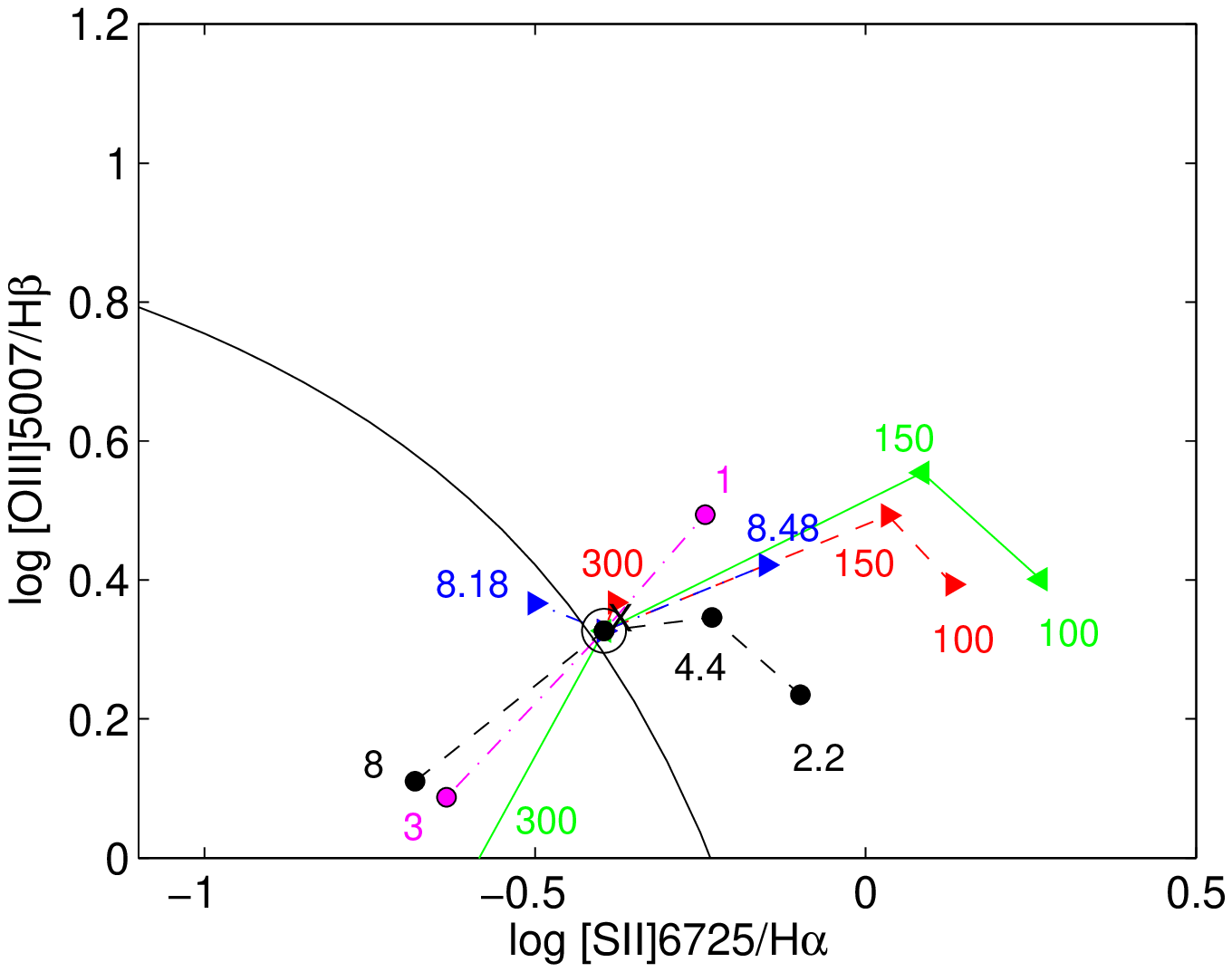} \\
\includegraphics[angle=0,width=1.0\columnwidth]{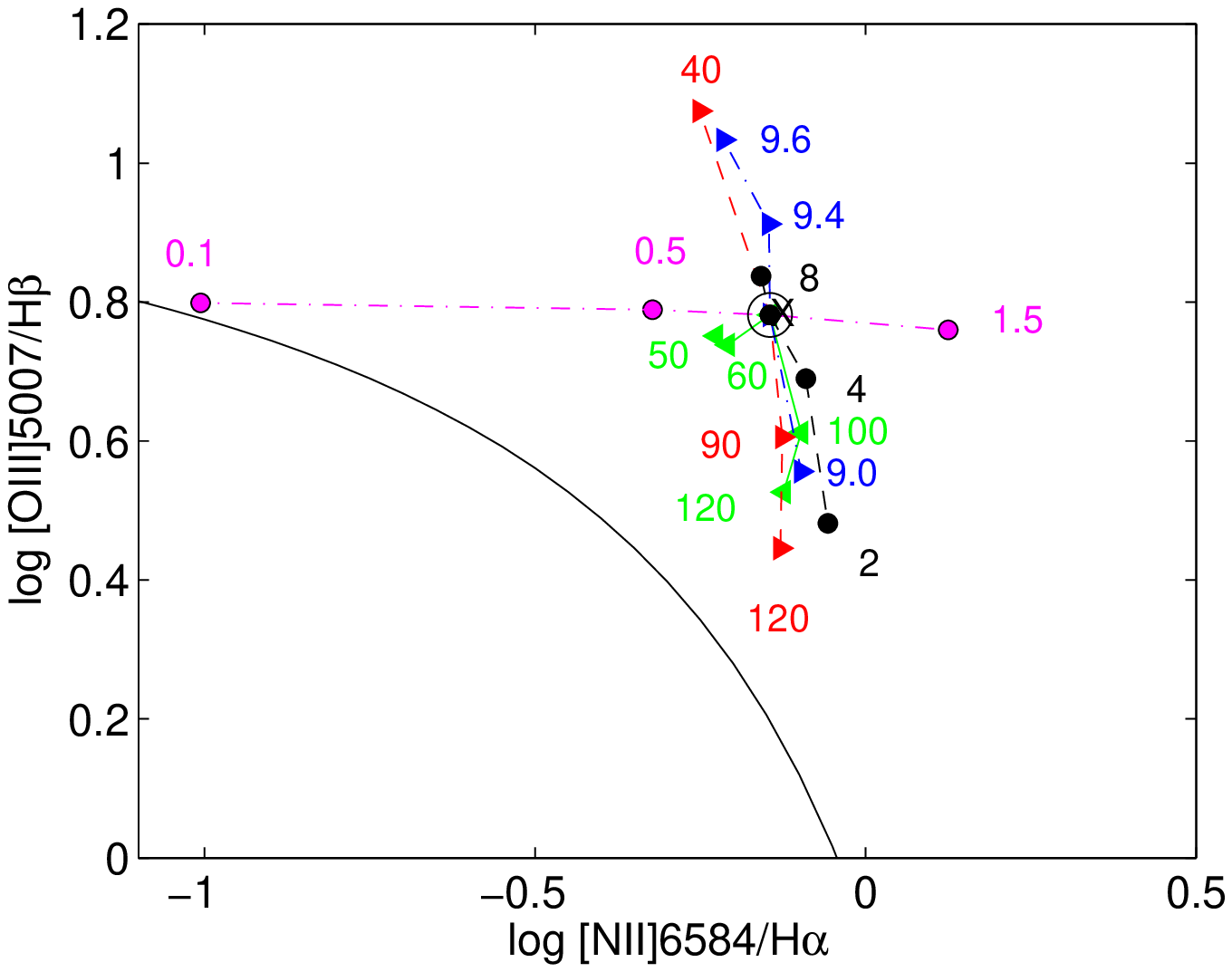} 
\includegraphics[angle=0,width=1.0\columnwidth]{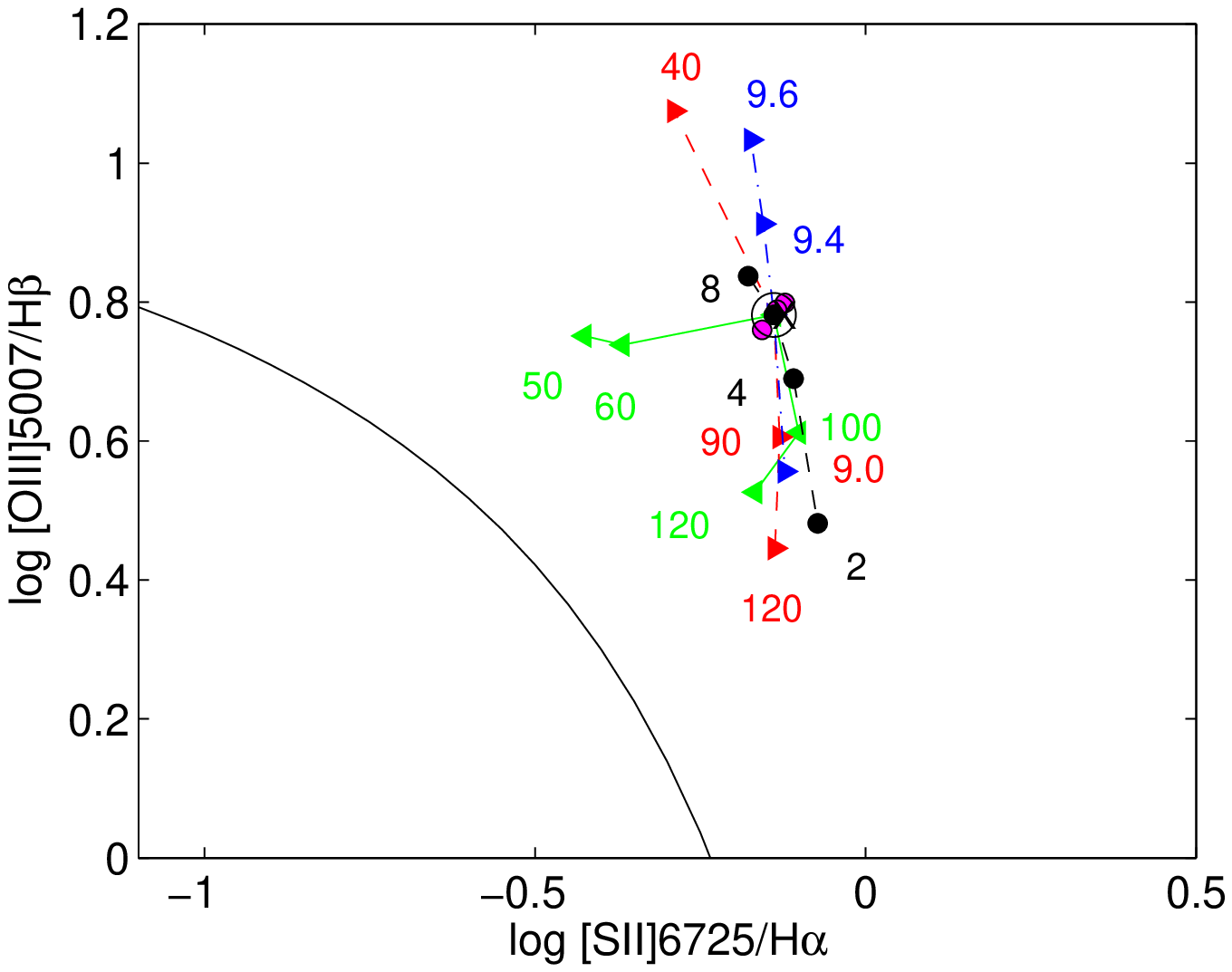}
\caption{Diagnostic diagrams showing the fit procedure for two objects  
  m31 (top panels) and m359 (bottom panels)
of our sample (see Sect.~\ref{obs}).   Curves represent the criteria proposed by \citet{kewley01} and \citet{enrique13} to separate
AGN-like from \ion{H}{ii}-like objects. The emission line [\ion{S}{ii}]$\lambda$6725 represents the sum of the intensities
of $\lambda$6716 and $\lambda$6731.
The lines show composite model predictions considering different input parameters of the
grid, i.e. for \Vs\, (green solid line), \n0\, (red dashed line), log($F$) (dot-dashed blue line),
 N/H ({\bf magenta} solid line) and O/H (black dashed line). The numbers near green triangles give the
 \Vs\, values in \kms, near red triangles the \n0 values in \cm3, near blue triangles the log($F$) values.
Black filled circles connected by a solid lines show the abundance ratios N/H (in 10$^{-4}$ units), those connected
by  dashed lines show the O/H abundance ratios (in 10$^{-4}$ units).
The black cross shows the observation data. The open black circle represents the model selected which best reproduces the data.}
\label{fitr1}
\end{figure*}

We calculate for each  object of our sample a large grid of models, varying the input
parameters \Vs, \n0, $D$, $F$, N/H, O/H,  and S/H,
in a consistent way (i.e. by considering the effect of each
of them on the different line ratios), until a fine tune of all  the
line ratios  to the data has been obtained. 
The best fit  models were finally selected by comparing the calculated to the  observed line
ratios  and by constraining the precision of the fit
by discrepancies that are set at 20 per cent for the strongest lines
(e.g. [\ion{O}{ii}]$\lambda$3727, [\ion{O}{iii}]$\lambda$5007) and 50 
per cent for the weakest lines (e.g. [\ion{O}{iii}]$\lambda$4363).
 We verify that the range of  parameters of the best fit  models
 do not differ  more than $10$ percent. Thus, this value is adopted as uncertainty in 
 the derived parameters from the composite models. 
  In order to show the fit procedure,
 in Figure.~\ref{fitr1}, two BPT diagnostic diagrams, we show a selected portion of the grids which were used to reproduce
 the spectra, for sample, of the  objects number 31 (\Vs$\rm \sim 200 \: km \: s^{-1}$)
 and 359 (\Vs$\rm \sim 80 \: km \: s^{-1}$). 
The best fitting model represented by the large open circle is slightly displaced from the observed
point (represented by a cross) in order to reproduce as much as possible all the line ratios
(see Table~\ref{tab1}). Not always the results follow smoothly and monotonically 
 the input parameters, in particular the element abundances. This is due to the fact that all of them participate
differently to the cooling rate throughout the cloud. Particularly, oxygen is a strong coolant.

In the present work, the   O/H, N/H and S/H relative abundances  were varied in the models in order
to  reproduce the correspondent observed lines. 
On the contrary, Ne/H and Ar/H  were kept constant in the models,
i.e. the solar values were  adopted for these elements. This has a  small affect
on the fitting models  as  they do not  dominate the cooling 
rate downstream as much as N, O and S.

 \citet{contini01}  presented a grid of composite models 
for narrow-line regions of active galaxies 
calculated with the SUMA code,  taking into account different values of shock velocities,
preshock densities, geometrical thickness of the clouds and
ionizing radiation intensities in a large range.
Based on this large grid,  they found that 
if the flux from the AC  is low ($F \rm \: \lid \: 10^{9} \: ph \: cm^{-2} \: s^{-1} \: eV^{-1})$,
a shock-dominated regime is found. \citet{contini01}  also showed that the [\ion{O}{iii}]/[\ion{O}{ii}]
line ratio  is much more sensitive to the intensity of
the flux radiation from the AC than to the shock velocities, being shock-dominated models
characterised by relatively
high  [([\ion{O}{ii}]$\lambda$3727/[\ion{O}{iii}]$\lambda$5007) $\ga \:1$] line ratios. 
The  ionization parameter $U$ can be obtained from the parameter $F$ by 
$U$= [$F$/($n \: c$ ($\alpha -1)$)]$\times$[($E_H)^{-\alpha +1}$$-$($E_C)^{-\alpha +1}]$
(see \citealt{contini83}), where
$E_H$ is hydrogen ionization potential  and $E_C$ is the high energy cutoff,
$n$ the density, $\alpha$ the spectral index, and $c$ the speed of light.

\begin{figure}
\centering
\includegraphics[angle=-90,width=1.0\columnwidth]{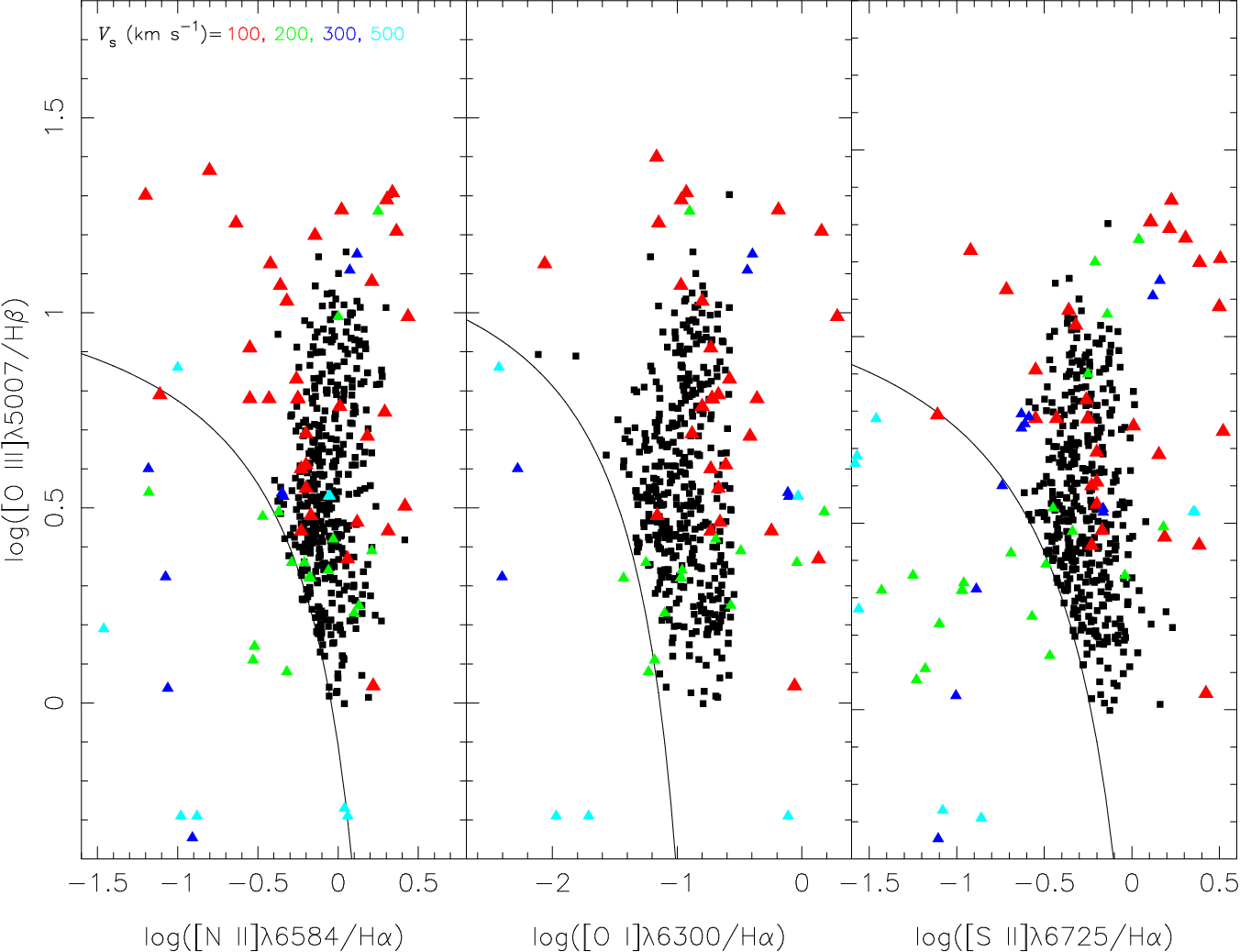} 
\caption{Diagnostic diagrams [\ion{O}{iii}]$\lambda$5007/H$\beta$
versus [\ion{N}{ii}]$\lambda$6584/H$\alpha$, versus [\ion{O}{i}]$\lambda$6300/H$\alpha$
and versus [\ion{S}{ii}]$\lambda$6725/H$\alpha$. Black points represent the 
463 objects of our sample. Colored points represent results of the composite model
grid built by \citet{contini01} and results  calculated 
for the objects of our sample m31 and m359, which include different values for N/H, O/H
and S/H, i.e. composite model results with a given range of  metallicity ($0.4 \: \la \:  (Z/Z_{\odot})\: \la \: 1.6$).
Result models with different
gas shock velocity (\Vs~, in units of $\rm km\: s^{-1}$) are plotted with different
colours, as indicated.
The lines represent the  criteria proposed by \citet{kewley01} and \citet{enrique13} to separate
AGN-like from \ion{H}{ii}-like objects.}
\label{figmr1}
\end{figure}

 In Fig.~\ref{figmr1}, we present three BPT diagrams containing the emission-line ratio intensities
of the observational sample and  those predicted by the grid of composite models built
by \citet{contini01}. 
The model predictions are discriminated in terms of
the shock velocity (\Vs\,, in $\rm km \: s^{-1}$) as indicated in the diagrams.  
Since the models of this grid consider only solar metallicity,
we have added the results calculated for the grids adopted for m31 and m359 which include different values for N/H, O/H
and S/H, i.e. composite model results with a range of  metallicity ($0.4 \: \la \:  (Z/Z_{\odot})\: \la \: 1.6$).
Fig.~\ref{figmr1} shows that the models cover very well the region occupied by the observations.

\subsection{Detailed model approach}

 Most of the studies carried out to derive metallicity or abundances in AGNs
 are either based on theoretical calibrations from photoionization model sequences
(e.g. \citealt{thaisa98, dors14}) or on comparisons  
between model predicted and observational line-intensity ratios in diagnostic diagrams 
(e.g. \citealt{nagao06a, feltre16, castro17, dors19, sarita20}).
The main problem in using photoionization model sequences is that, in most cases, it is assumed
fixed relations for  the N-O and S-O abundances, which can produce
very uncertain abundance results \citep{enrique09}, mainly because after the oxygen, 
the nitrogen and the sulphur are the main coolant elements in the nebular gas. 
Moreover, the above abundance relations are poorly known in AGNs (see \citealt{dors17, dors19}).

In this sense, the use of detailed modeling (e.g. \citealt{dors17, contini17}) or  
bayesian-like comparisons (e.g.\ \citealt{enrique19})
circumvent this problem  because not fixed relations between the N, S and O abundances are assumed, producing a more accurate solution for the nebular  thermal equilibrium and, consequently, more reliable abundances.
Although in the detailed models used in this work  a fixed slope for the power law
representating the SED and a fixed dust-to-gas ratio were assumed, \citet{feltre16} showed that these parameters
have a secondary influence on the model predicted emission-lines.

\begin{figure*}
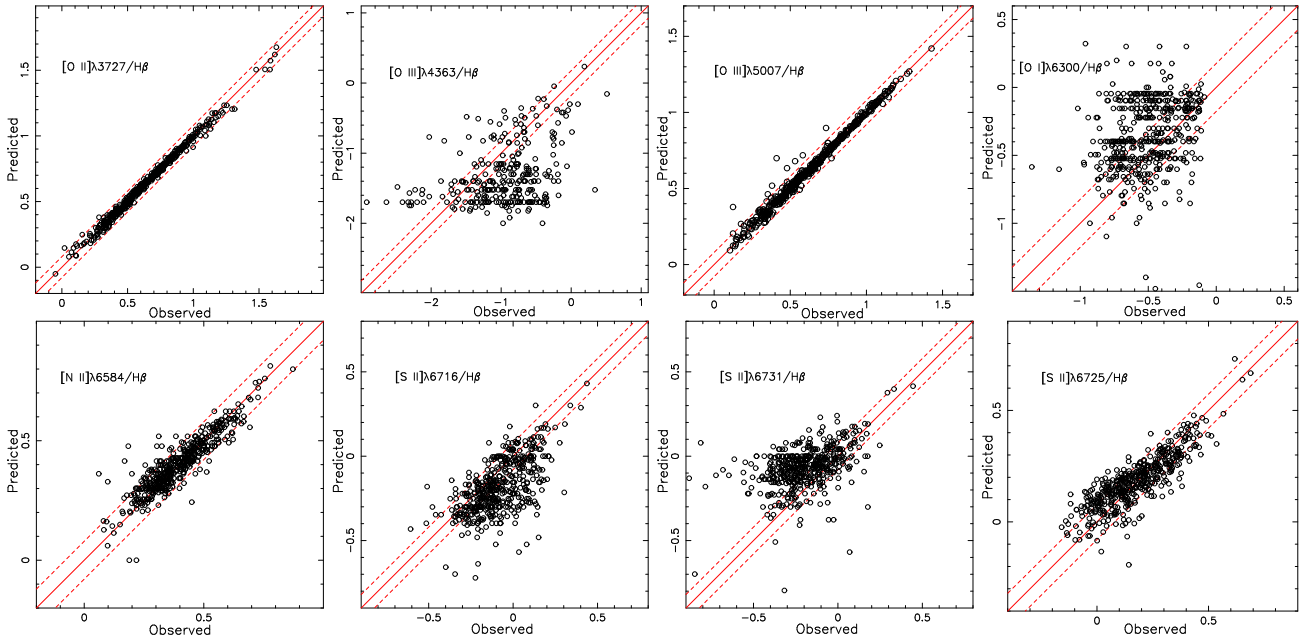

\centering
\includegraphics[angle=-90,width=0.5\columnwidth]{comp_3727c.eps} 
\includegraphics[angle=-90,width=0.5\columnwidth]{comp_4363c.eps}
\includegraphics[angle=-90,width=0.5\columnwidth]{comp_5007c.eps}
\includegraphics[angle=-90,width=0.5\columnwidth]{comp_6300c.eps}
\includegraphics[angle=-90,width=0.5\columnwidth]{comp_6584c.eps}
\includegraphics[angle=-90,width=0.5\columnwidth]{comp_6716c.eps}
\includegraphics[angle=-90,width=0.5\columnwidth]{comp_6731c.eps}
\includegraphics[angle=-90,width=0.5\columnwidth]{comp_6725c.eps}
\caption{Comparison between the logarithm of model predicted and observed emission-line fluxes relative to the H$\beta$ flux for the sample of
 463 Sy2s
(see Section~\ref{obs}). Solid lines represent the one-to-one relation. Dashed lines  show the deviation,
representing the  observational uncertainty,  of the equality by a factor 0.1 dex and 0.2 dex for strong and weak
emission lines, respectively. The line   [\ion{S}{ii}]$\lambda$6725 corresponds to the sum
	of the emission-lines [\ion{S}{ii}]$\lambda$6716 and [\ion{S}{ii}]$\lambda$6731. }
\label{fig1}
\end{figure*}

 \begin{figure}
\centering
\includegraphics[angle=-90,width=1.0\columnwidth]{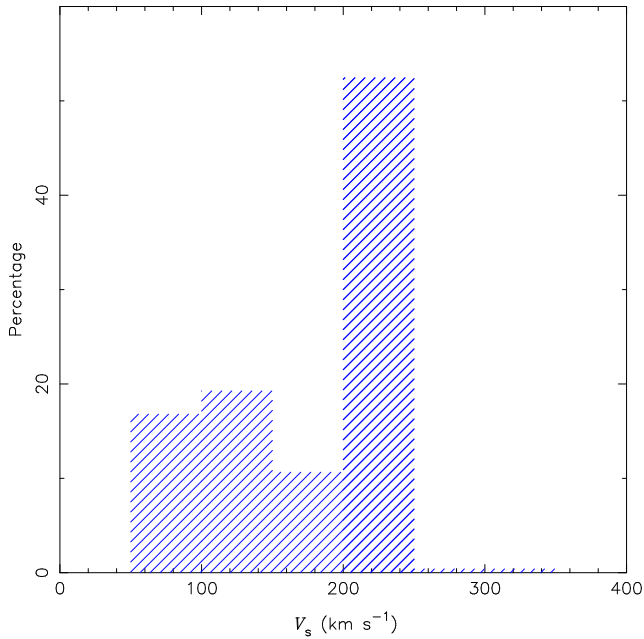} 
\caption{Distribution of shock the velocities \Vs~ 
predicted by the composite models (see Sect.~\ref{model}) for our sample, in velocity bins of 50 km $\rm s^{-1}$.}
\label{fig2}
\end{figure}


\section{Results}
\label{res}

In Figure~\ref{fig1}, the observed emission-line  ratios  are compared with those predicted
by detailed modelling  with {\sc SUMA} for the 463 objects.
The [\ion{O}{iii}]$\lambda$4363 line was measured only in 280/463 objects. 
In this figure  we also  show as dashed lines a typical observational uncertainty of 20 and 50 per cent
for strong and weak emission-line  ratios, respectively (e.g. \citealt{kraemer94}). 
A good agreement between the strong  [\ion{O}{ii}]$\lambda$3727,
[\ion{O}{iii}]$\lambda$5007 and  [\ion{N}{ii}]$\lambda$6584 emission-lines can be seen. 
The sum of the [\ion{S}{ii}]$\lambda$6716 and $\lambda$6731 lines is well reproduced by the models.

From the 463 objects of our sample, we obtain reliable model solutions for all the  strong emission 
lines [\ion{O}{ii}]$\lambda$3727, [\ion{O}{iii}]$\lambda$5007, [\ion{N}{ii}]$\lambda$6584,
and [\ion{S}{ii}]$\lambda$6716+$\lambda$6725 in 244 galaxies.
In Table~\ref{tab1} the observed intensity ratios are compared with the  calculated ones (relative to \Hb=1),
while  the model parameters selected  from the best fit to the observed data  are listed in Table~\ref{tab2}.
The results presented in Table 1 show some major discrepancies  for the [\ion{O}{iii}]$\lambda$4363/\Hb, 
([\ion{S}{ii}]$\lambda$6716,$\lambda$6731)/H$\beta$ 
and [\ion{O}{i}]$\lambda$6300/\Hb~ line ratios  in about half of the observed spectra.
The [\ion{O}{iii}]$\lambda$4363 line is strongly blended with the \Hg$\lambda$4340 line, in particular for shock 
velocities $\geq$ 100 \kms. Therefore, the results for the calculated [\ion{O}{iii}]$\lambda$4363/\Hb~ line ratios
 can differ from  the values presenting contamination by the \Hg~ line.
Moreover, some of the  observed  [\ion{S}{ii}]$\lambda$6716/$\lambda$6731 line ratios are $>\: 1$, while the calculated 
ones are $< \:1$  
in some objects  (Table 1) and for only  57 objects   the models were able to 
reproduce the  [\ion{O}{i}]$\lambda$6300/H$\beta$ observational line ratio,
with a difference smaller than 50 per cent between the observed and predicted intensities.
These problems were  explained by \citet{congiu17}, who pointed out that the contribution of the
interstellar medium (ISM)  to the  extended NLR is particularly high for the [\ion{S}{ii}] and [\ion{O}{i}] lines.
These lines are emitted  by relatively low temperature gas ($T_{\rm e}\:\leq \:10^{4}$ K).
 The first ionization potential of sulphur is lower than that of H and  
the  oxygen first ionization potential   is similar to that of H.
Therefore, these lines can be relatively strong in the ISM as well as those
 emitted  from an ionized-neutral transition zone, located in outskirt layers of the nebulae.
Moreover, the [\ion{S}{iii}]$\lambda$6312 line  can be blended with the 
[\ion{O}{i}]$\lambda$6300, $\lambda$6363 doublet, leading to  further
discrepancies  in reproducing the [\ion{O}{i}] lines by the models.

\begin{figure} 
\includegraphics[scale=0.6, angle=-90]{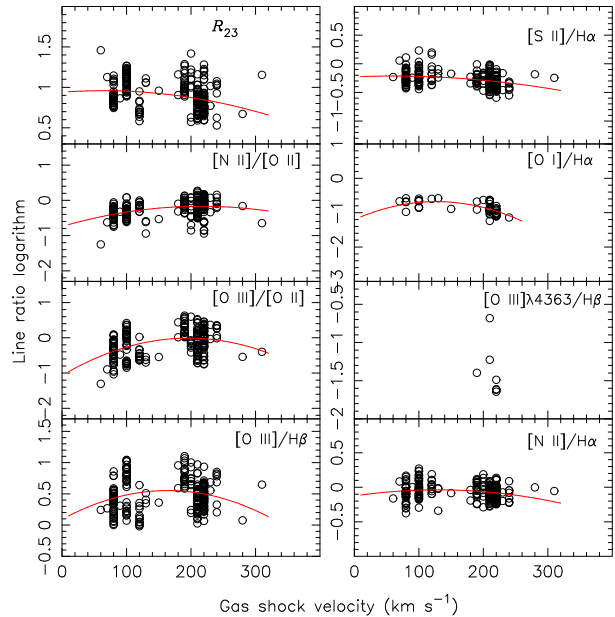}
\caption{Observed line ratios  of the sample  objects (see Section~\ref{obs}) 
versus the gas shock velocity predicted by the  models (see Sect.~\ref{model}). In each panel
the related  line ratio  is indicated. Curves represent the fit of a  second order polynomial whose
coefficients are listed in Table.~\ref{tab3}.}
\label{fig3}
\end{figure}

\begin{table*}\small
\caption{De-reddened observed (Obs.) and predicted (Mod.)  fluxes (relative to H$\beta$=1.00) 
 for two of the objects in our sample of Seyfert 2 nuclei. 
The observed values  are  listed in the lines with the identification
of each object. The  predicted values
 are listed in the lines starting with the label  m.
 The stellar mass and redshift of each object are listed in the full table, available as supplementary material.
 In cases which a line was not measured its value is refered to 0.00.}
\label{tab1}
\begin{tabular}{lcccccccccccccc}	 
\noalign{\smallskip} 
\hline 
    &     Object            & [\ion{O}{ii}] & [\ion{Ne}{iii}] & [\ion{O}{iii}] &  [\ion{O}{iii}]               &  \ion{He}{i} & [\ion{O}{i}]  &  H$\alpha$	  &[\ion{N}{ii}]  &[\ion{S}{ii}]   & [\ion{S}{ii}]  &[\ion{Ar}{iii}] \\ 
    &                       & $\lambda$3727 & $\lambda$3869   & $\lambda$4363  & $\lambda$4959+$\lambda$5007   & $\lambda$5876 & $\lambda$6300 & $\lambda$6563 & $\lambda$6584 & $\lambda$6716  &  $\lambda$6731 & $\lambda$7135  \\
3   &	J000819.72-000002.7 &  2.58 	    & 0.00   	      & 0.00 	       & 2.77		               &  0.00	 & 0.27          & 2.86      	  & 2.37 	 & 0.84 	  & 0.35	   & 0.14	     \\
m3  &            ---        &  2.90 	    & 0.75  	      & 0.45 	       & 2.76		               &  0.10	 & 0.30          & 3.07      	  & 2.20 	 & 0.43 	  & 0.75	   & 0.10	     \\
4   &	J000908.27-011013.8 & 1.64  	    & 0.73  	      & 0.00 	       & 2.13		               &  0.01	 & 0.31          & 2.86      	  & 2.37 	 & 0.78 	  & 0.27	   & 0.00	     \\
m4  &            ---        & 1.52  	    & 0.51  	      & 0.45 	       & 2.00		               &  0.13	 & 0.67          & 3.05      	  & 2.50 	 & 0.50 	  & 0.80	   & 0.83	     \\
\hline
\end{tabular}
\end{table*}

\begin{table*}\small
\caption{Model parameters selected to fit the observation emission-line ratios
of each object of the sample.  Shock velocity \Vs~ in units of $\rm km \: s^{-1}$.
Atomic pre-shock density \n0 in units of $\rm cm^{-3}$. 
Geometrical thickness of the cloud ($D$) in units of $10^{16}$ cm.
Radiation flux ($F$)  emitted by the AGN (primary source) in units of $10^{10}$  photons $\rm cm^{-2}\:  s^{-1} \: eV^{-1}$ at the Lyman limit,
 measured at the inner surface of the cloud.
Abundances of N/H, O/H and S/H are in units of $10^{-4}$. The  flux of H$\beta$ [$F$(H$\beta$)] in  units 
of $\rm erg \: cm^{-2} \:  s^{-1}$ is calculated at the nebula.
Full table is available as supplementary material.}
\label{tab2}
\begin{tabular}{lcccccccc}	 
\noalign{\smallskip} 
\hline   
Model   &   \Vs   &   \n0  &      $D$   &     $F$   &   N/H    &    O/H      &   S/H   &    $F$(H$\beta$) \\
m3      &    220  &  260   &      7.2   &     1.2   &     1.0  &    6.6      &    0.1  &     0.071	   \\
m4      &    240  &  300   &      9.2   &     2.0   &     1.2  &    6.6      &    0.2  &     0.180	   \\
\hline
\end{tabular}
\end{table*}

In Fig.~\ref{fig2}, a  histogram  with the distribution of shock velocities \Vs\, (in units of $\rm km \: s^{-1}$)
predicted by the  models for our sample is shown.
We find that the clouds in the NLR  have \Vs\, ranging from 60 to 310 $\rm km \: s^{-1}$, with an 
average value of $\sim 170$ $\rm km \: s^{-1}$. For  half of the objects  ($\sim 52$ per cent)
\Vs~  ranges between  200 \kms and 250 $\rm km \: s^{-1}$. Moreover,  
in Fig.~\ref{fig3}, the  values of some
observed line ratios in our sample versus \Vs\, and the resulting polynomial fits
of the points, represented by  curves, are shown. Most of these line ratios were  considered by \citet{dors20a}
to be used in metallicity estimations of AGNs.
The fitting coefficients are listed in Table~\ref{tab3}. 
The [\ion{O}{i}]$\lambda6300$/H$\alpha$ and [\ion{O}{iii}]$\lambda4363$/H$\beta$ line 
ratios are shown only for the cases in which the models were able to reproduce them, taking into account
the uncertainty of 50 per cent.

It can be noticed that:
\begin{enumerate}
\item the  [\ion{O}{iii}]$\lambda5007$/H$\beta$ and [\ion{O}{iii}]$\lambda5007$/[\ion{O}{ii}]$\lambda3725$
line ratios increase with \Vs\, until  $\rm \sim 150 \: km \: s^{-1}$ and decrease
for higher velocity values. 
\item  $R_{23}$  is approximately constant for $V_{\rm s} \: \la \rm \: 150 \: km \: s^{-1}$ and it decreases
for higher velocities.
\item The [\ion{N}{ii}]$\lambda6584$/[\ion{O}{ii}]$\lambda3727$, [\ion{N}{ii}]$\lambda6584$/H$\alpha$,
[\ion{O}{i}]$\lambda6300$/H$\alpha$, and [\ion{S}{ii}]$\lambda6725$/H$\alpha$ line ratios are approximately constant
in the range of the \Vs\,  derived values.
\item  No conclusions can be obtained from [\ion{O}{iii}]$\lambda4363$/H$\beta$ versus \Vs\, due to
the small number of points.
\end{enumerate} 
 It is worth  mentioning that, despite the fitting coefficients (see Table~\ref{tab3}) indicate that there
seems to be a correlation betweew the line ratios  and \Vs, the above result is extremely marginal, 
 due to the scattering of the points
at a fixed velocity is large. Detailed modelling of a set of lines observed in distinct observational ranges (e.g. optical and ultraviolet)
could put more constraints to the models and produce more reliable results, 
confirming the trend observed in Fig.~\ref{fig3}.

The observed behaviour of \Vs~ with the emission-line ratios involving the oxygen lines 
(e.g. [\ion{O}{iii}]/[\ion{O}{ii}]) is due to the increase of  $T_{\rm e}$ ($\propto V_{\rm s}^{2}$)  that results  in an increase of 
 the intensities of the lines more sensitive to this parameter,
 i.e. [\ion{O}{iii}]$\lambda$5007 (and eventually oxygen lines from higher ionization levels)
 rather than [\ion{O}{ii}]$\lambda$3727.
However, for $V_{\rm s} \: \ga \: 150 \: \rm km \:s^{-1}$, 
the considerable increase of  the ionization degree results in an increase  of the $\rm O^{3+}$ ion  abundance, 
decreasing  the  [\ion{O}{iii}]$\lambda5007$ emission.  The nitrogen and sulphur lines are less sensitive to \Vs. 
The gas density increases with \Vs, therefore  the decrease of [\ion{N}{ii}], [\ion{O}{ii}] and [\ion{S}{ii}] lines 
is mostly due to their relatively low critical density for collisional de-excitation.

\begin{figure*}
\centering
\includegraphics[width=8.8cm]{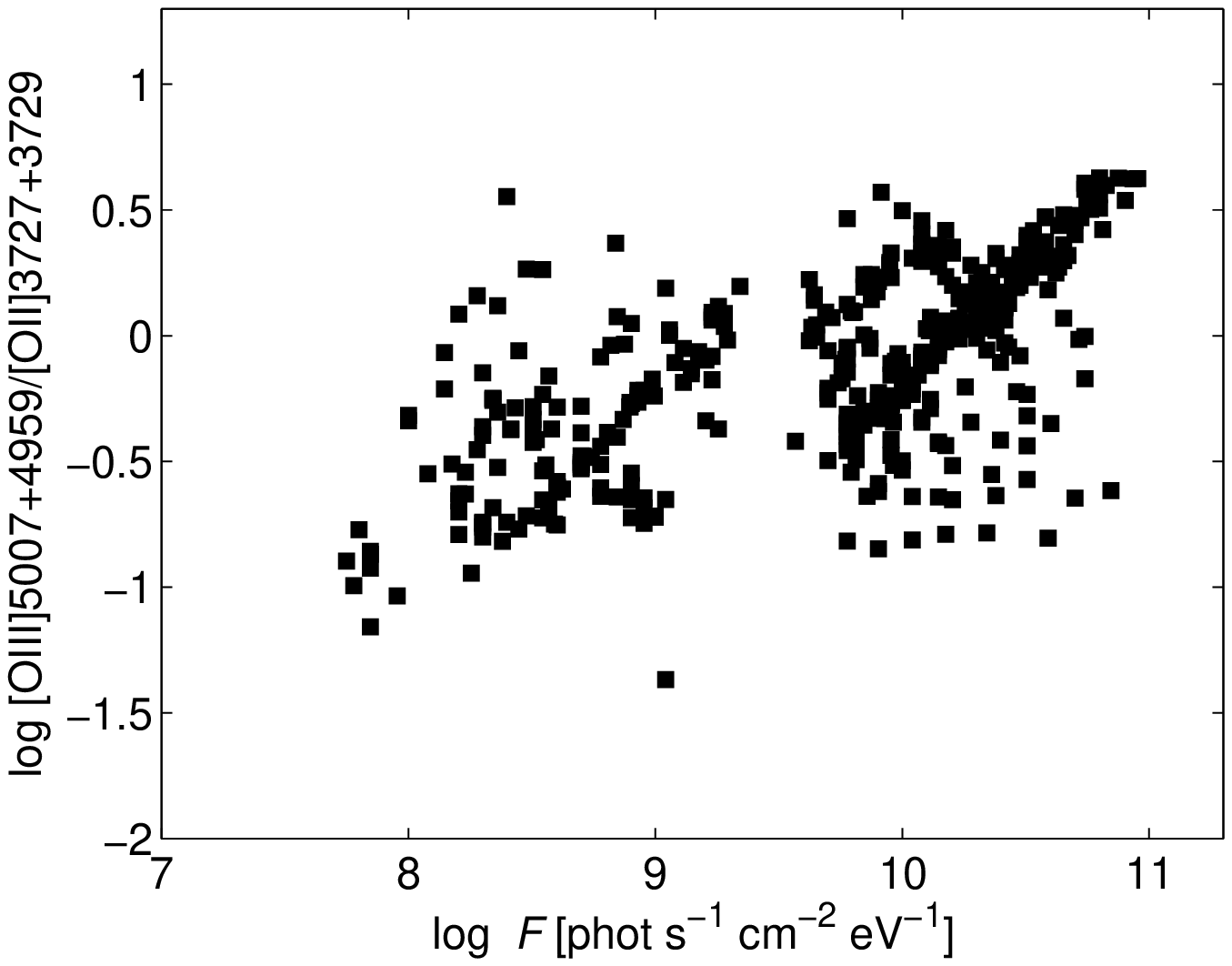}
\includegraphics[width=8.8cm]{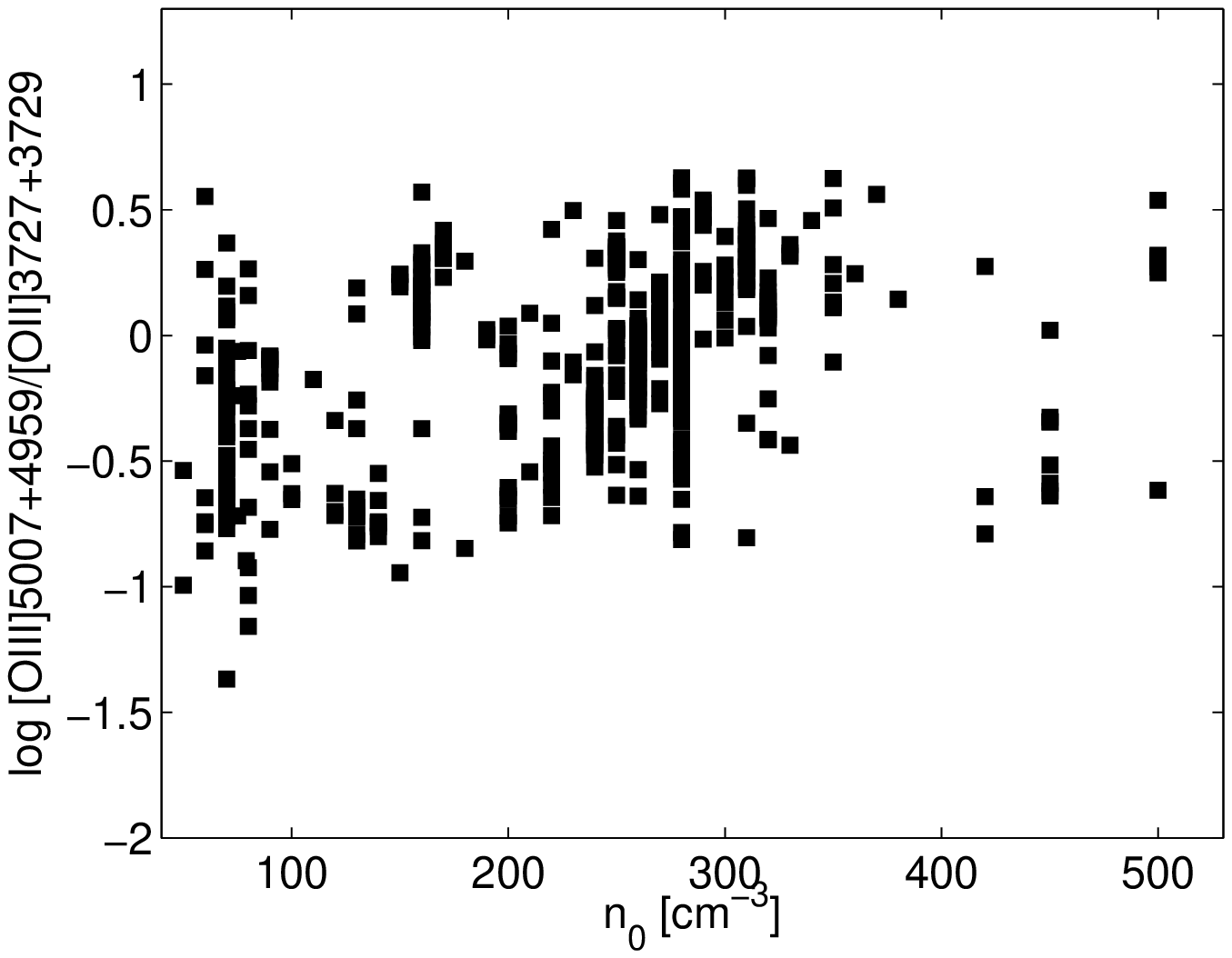}
\caption{Diagrams  of [\ion{O}{iii}]($\lambda$5007$+\lambda$4959)/[\ion{O}{ii}]$\lambda$3727 intensity ratios 
versus the logarithm of the flux radiation ($\log F$) from the AC (left panel)
and pre-shock density (\n0, rigth panel). }
\label{discr1}
\end{figure*}

\begin{figure*}
\includegraphics[width=8.8cm]{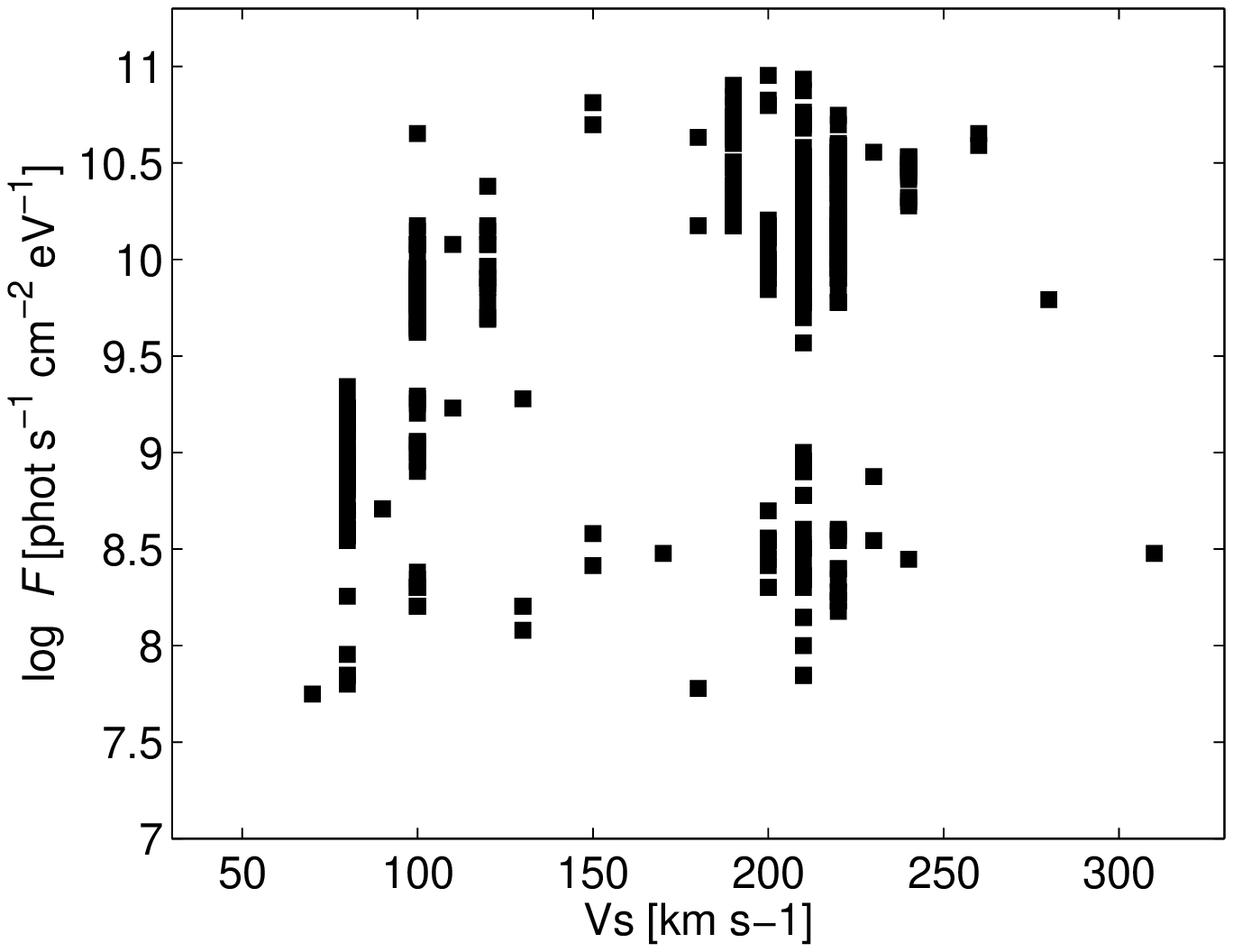}
\includegraphics[width=8.8cm]{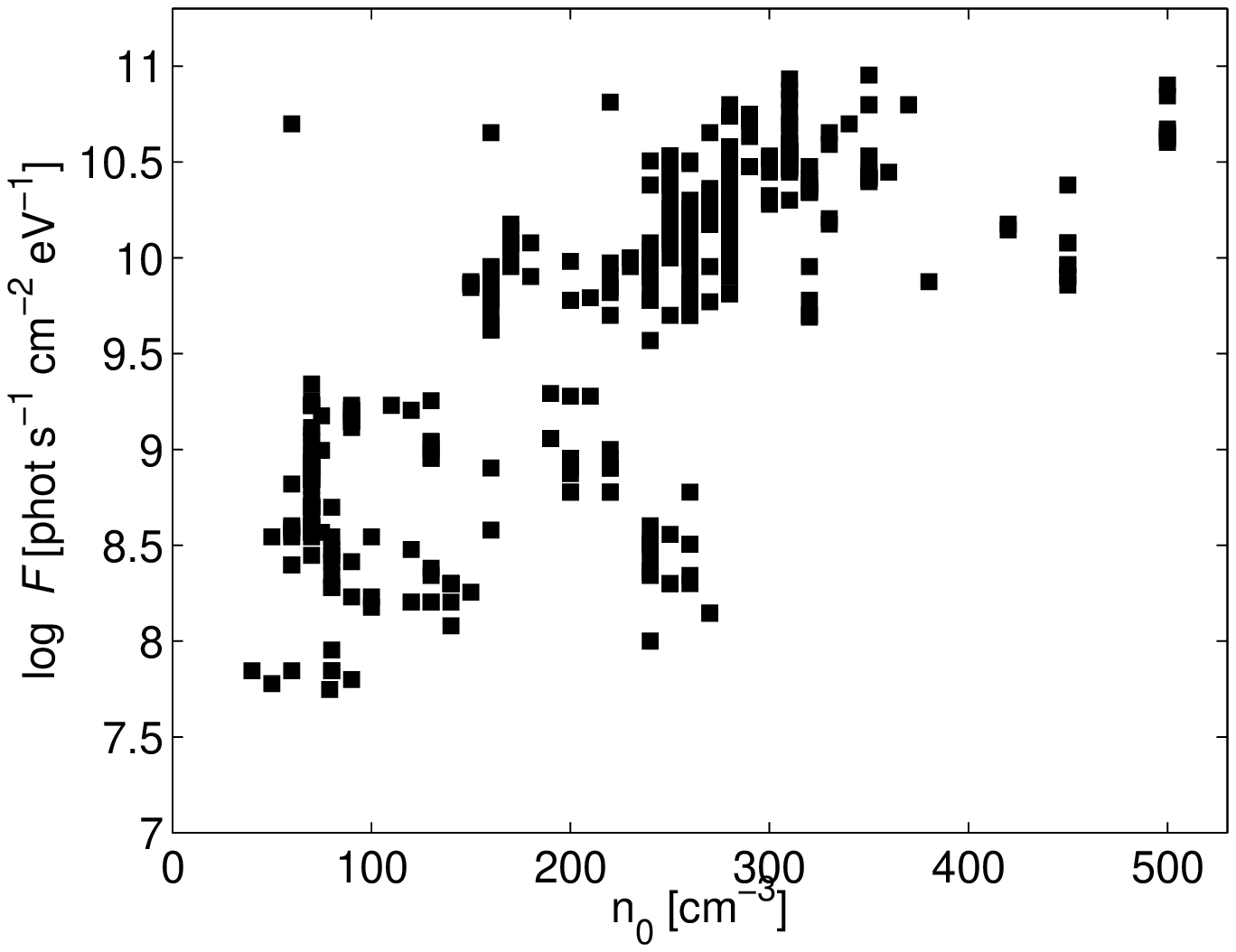}
\caption{Logarithm of the flux radiation $\log F$ from the AC versus the shock velocity (\Vs, left panel)
and versus the pre-shock density (\n0, rigth panel).}
\label{anvsdic2}
\end{figure*}

One can also note in  Figs.~\ref{fig2} and \ref{fig3} that the resulting distribution of velocities \Vs\, 
in the models is not
continuous, with the points being grouped more or less around velocities of 100 and 200 $\rm km \: s^{-1}$
 and few points in between these values. We present an additional analysis in order to explain this results.
 Firstly, our sample of Seyfert~2 galaxies is rich enough in number of objects  to confirm some results obtained
 for AGNs in general and the result above can not be associated to the sampling of the models.
In Fig~\ref{discr1}, diagrams of our composite model predictions
for [\ion{O}{iii}]$\lambda$5007/[\ion{O}{ii}]$\lambda$3727
line ratios  versus  flux radiation ($F$) from the AC  and versus the  pre-shock density
(\n0)  are presented in order to understand the role of the main physical  parameters.
In the left panel of Fig.~\ref{discr1} it can be seen
that the galaxies are grouped around $\log F \: \la  \: 9.3$,
$\log F \: \ga \: 9.6$   and 
there seems to be a gap in the range of $9.3 \: \la \: \log F \:  \la \: 9.6$ 
($F$ is in units of $\rm ph \: cm^{-2} \: s^{-1} \: eV^{-1})$.
However, in right panel of Fig~\ref{discr1},  
a continuum distribution between [\ion{O}{iii}]/[\ion{O}{ii}] and \n0
is derived.  Another result is shown in Fig.~\ref{anvsdic2},  
where $F$ is plotted as a function of the shock velocity (left panel) and 
pre-shock density  (right  panel). 
Clearly, it is possible to see few points around
$\sim$150 \kms confirming the dichotomy and, again, a continuum behaviour of  $n_{0}$ 
is found and, obviously,  the gap in $F$ is also present.
 We suggest that objects with $9.3 \: \la \: \log F \:  \la \: 9.6$ or with \Vs\, around $\sim 150$ \kms
could probably correspond to LINERs (see  \citealt{contini97} and references therein), considering 
that lower velocities would correspond to SF regions and higher velocities to Seyfert galaxies. 
As LINERs were excluded from our sample  due to the criteria adopted by \citet{dors20a}, 
this could explain the gap observed in the shock velocities.

\begin{table}\small
\caption{Coefficients of the fitting of $R={\rm a} V_{\rm s}^{2} + {\rm b} V_{\rm s} + {\rm c}$ to the
points shown in Fig.~\ref{fig1}. $R$ correspond to different line ratios as indicated.}
\label{tab3}
\begin{tabular}{lccr}	 
\noalign{\smallskip} 
\hline   
 $R$                                 &  a ($\times10^{-5}$)&  b ($\times10^{-3}$)  & \multicolumn{1}{c}{c}\\			      
$[$\ion{O}{iii}$]$/H$\beta$          & $-1.72(\pm0.75)$	   &  $5.64(\pm2.35)$	   & $+0.09(\pm0.16)$ \\   				     
$[$\ion{O}{iii}$]$/$[$\ion{O}{ii}$]$ & $-2.76(\pm1.00)$	   &  $10.82(\pm3.15)$	   & $-1.07(\pm0.21)$ \\  					     
$[$\ion{N}{ii}$]$/$[$\ion{O}{ii}$]$  & $-1.23(\pm0.65)$	   &  $5.30(\pm2.03)$	   & $-0.74(\pm0.14)$ \\  					     
$R_{23}$                             & $-0.46(\pm0.51)$	   &  $0.59(\pm1.59)$	   & $+0.94(\pm0.10)$ \\   				     
$[$\ion{N}{ii}$]$/H$\alpha$          & $-0.52(\pm0.33)$	   &  $1.36(\pm1.04)$	   & $-0.12(\pm0.07)$ \\  					     
$[$\ion{O}{i}$]$/H$\alpha$           & $-3.20(\pm1.25)$	   &  $8.17(\pm3.86)$	   & $+1.05(\pm0.10)$ \\   				     
$[$\ion{S}{ii}$]$/H$\alpha$          & $-0.36(\pm0.36)$	   &  $0.42(\pm1.13)$	   & $-0.12(\pm0.07)$ \\  					     
\hline
\end{tabular}
\end{table}


Concerning the results for the element abundances of our sample, the models predict oxygen abundances
in the range $\rm 8.5 \: \la \: [12+\log(O/H)] \: \la \: 8.9$, with an average value 
of  $8.8\pm 0.03$. Adopting the solar value  $12+\log(\rm O/H)_{\odot}=8.89$ \citep{allendeprieto01},
the  O/H values above correspond  to  the metallicity range  
$0.6 \: \la \:  (Z/{\rm Z_{\odot}}) \: \la \: 1.6$, and an average value
$(Z/{\rm Z_{\odot}})\approx 1.3$. The logarithm of the N/O abundance ratio is  in the
range $\rm -1.1 \: \la \: [\log(N/O)] \: \la \: -0.3$ with an average value  $-0.8\pm 0.12$.


\section{Discussion}
\label{disc}

Shocks  created by the interaction of radio jets/outflows with
the surrounding ISM, in addition to photoionization by  radiation  from the accretion disk, have a strong influence on  
the observed  emission lines of AGNs.
Observations carried out along decades have shown that  outflows are  commonly observed in 
AGNs (for a review see, e.g. \citealt{king15, harrison18}). 
In the early years,  outflows were observed mainly in powerful radio galaxies as, for instance,
in  3C405 (Cygnus A) by \citet{tadhunter91}, showing velocities around 1800 $\rm km \: s^{-1}$ 
and recent studies have shown that around  40 percent of the quasars  present outflows
(see \citealt{arav20} and references therein).  Concerning lower luminosity AGNs, such as Seyferts, 
outflows have also been observed, but with relatively lower velocities. For example, 
\citet{may18} found ionized outflows with velocities of $\sim 700 \: \rm km\: s^{-1}$
in the central 170 pc of the nearby Seyfert  nucleus ESO 428-G14. Using the  data from the MaNGA survey \citep{blanton17}, 
\citet{ilha19} studied the gas kinematics of a sample of 62 Seyferts and LINERs.
 By comparing their AGN sample with a sample of non-active galaxies  (see also, e.g. \citealt{riffel20, wylezalek20})
they found  that outflow signatures in the ionized gas within the central 1-2 kpc 
with velocity $v \: < \: 400 \: \rm km\: s^{-1}$ are seen in most AGN hosts. 

From a  theoretical point of view,  pure shock models (e.g. \citealt{binette85, dopita96, allen08, alarie19}) 
and composite
models (shock+AGN, e.g. \citealt{contini83,  contini86, congiu17, contini17, contini19}) have predicted
 shock velocities in Seyfert NLR   between 100 \kms and 500 $\rm km \: s^{-1}$.  From  the detailed modelling
of the present  large sample of Seyfert~2, we derived a similar range of shock velocities
(from 60 to 310 $\rm km \: s^{-1}$)  to those
 obtained in previous works. Moreover, this range of velocities is in agreement with those
found in  observational  investigations  of Seyfert 2 NLR,   strengthening  the confidence in our results.  
The present  modelling  of the  spectra, which  makes use of  a large range of  physical parameters, 
offers an unique opportunity
to investigate several properties of Seyfert galaxies, such as their position in diagnostic diagrams
as a function of different shock parameters, the temperature and ionization structure 
and chemical abundance determinations. Each of these  issues is  discussed in the following.

\subsection{Diagnostic Diagrams}

\citet{baldwin81} proposed that a combination of two pairs of line ratios, originally [\ion{O}{iii}]$\lambda$5007/H$\beta$
versus [\ion{N}{ii}]$\lambda$6584/H$\alpha$, can be used to discriminate the ionizing source of line emitting objects,
i.e. objects ionized by hot stars and by  a non-thermal source. These and other combinations of line ratios
are known as the BPT diagrams (see also \citealt{veilleux87, kewley01, kewley13, kauffmann03, enrique13, ji20a}). 
In particular, the sequence formed by AGN emission-line ratios
in BPT diagrams has been explained by variations in  the physical parameters  according to different
assumptions (e.g. \citealt{ji20}). 
\begin{enumerate}
\item Pure photoionization: considering photoionization of  AGN  clouds  by radiation from  gas acretion into  
a black hole, whose  SED can be represented by a power law. 
For example, \citet{feltre16} showed that the increase of certain line ratios, such as 
[\ion{O}{iii}]$\lambda$5007/H$\beta$, is mainly due to the combination of metallicity  with the 
ionization parameter
(see also \citealt{groves06}). A secondary dependence between the hardness of the SED and the electron density with 
the line ratios is also found (e.g. \citealt{feltre16, sarita20}). 
\item  Simple Equilibrium Model: \citet{flury20}, assuming  an approach for estimating  abundances of heavy elements, which 
involves a reverse-engineering of the direct method\footnote{Direct method or $T_{\rm e}$-method is based
on  elemental abundance of heavy elements calculated by using direct estimations of the electron temperature and
electron density.}, showed that high [\ion{O}{iii}]$\lambda$5007/H$\beta$
and [\ion{N}{ii}]$\lambda$6584/H$\alpha$ values are associated to high O/H   and ionization degree values of
the gas phase.
\item Pure shock heating and ionization effects: \citet{dopita95}, by using  
radiative steady-flow shock models \citep{dopita96}, showed that,
at a fixed solar metallicity, most of the line ratios involved in the BPT diagrams 
increase with \Vs\, (see also \citealt{allen08}).
\item Starburst-AGN mixing:  \citet{davies14a, davies14b}, who combined  stellar evolutionary synthesis models
from the {\sc Starburst99} code \citep{leitherer99} with the {\sc MAPPINGS}  photoionization code \citep{binette85},  
showed that for spatially resolved objects, hence the metallicity and the ionization parameters are fixed,
the increase of some line ratios in BPT diagrams can be interpreted in terms of 
the ionization flux fraction of  SFs to  AGN.
AGNs with the highest line intensity ratios have less  SF flux contamination.
\citet{thomas18} also considered a Starburst-AGN mixing  to interpret  SDSS data of AGNs and
found that, even for strong AGNs [with  log([\ion{O}{iii}]$\lambda$5007/H$\beta$) $\ga \:0.9$],  $\sim30$ per cent 
of the  Balmer line flux on average comes from \ion{H}{ii} regions.  
\end{enumerate} 

\begin{figure} 
\includegraphics[scale=0.6, angle=-90]{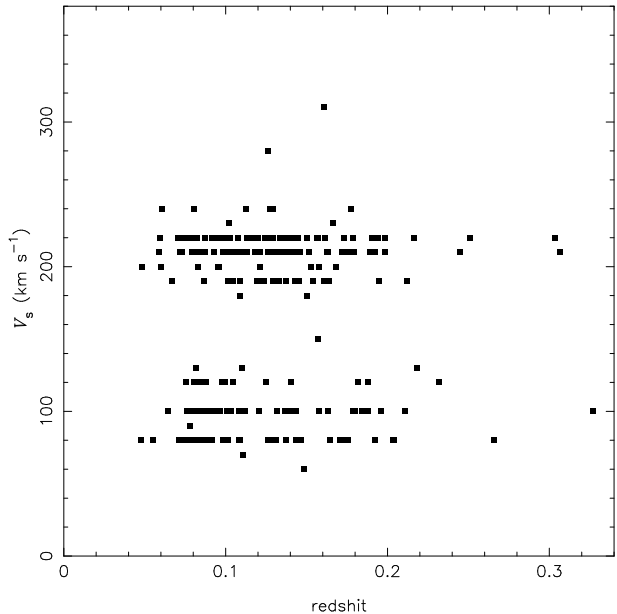}
\caption{Shock velocity \Vs~ predicted by the composite models versus the redshift 
for the sample objects.}
\label{fig6}
\end{figure}

Our sample is based on integrated SDSS spectra taken with a fixed optical fiber diameter of $\sim3$ arcsec, which 
corresponds to   a  physical scale (D) at the galaxies in the range $\rm 1.8 \: \la \: D(kpc) \: \la \:   15$. 
Although an   \ion{H}{ii} region contribution
is expected \citep{thomas18}, the   emission  from the sample galaxies is mainly from the AGN. In fact, \citet{dors20a}
did not find any correlation between the oxygen abundance and the electron density of AGNs with
the redshift, indicating that the  aperture effect does not affect the derived
parameters of the SDSS sample, at least for $z\: \la \:0.4$ (see also \citealt{kewley05}). As an additional test, in Fig.~\ref{fig6}, 
the  shock velocity derived from modelling
is plotted against the redshift value of each object of our sample. Since the 
contribution of \ion{H}{ii} region fluxes to the AGN tends to increase with the distance from the nuclei \citep{davies14a, davies14b} and, in general,
SFs  show low  shock velocities, a decrease of \Vs\, with $z$ would be expected if a significant fraction of the 
nuclear emission of the galaxies 
of our sample can be attributed to the contamination from extra-nuclear \ion{H}{ii} regions. However,  Fig.~\ref{fig6} 
shows no correlation between  \Vs\ and the redshift.
Therefore, we conclude that the  SFs flux contribution to the AGN spectra of our sample  is negligible and,
consequently,  the derived parameters based on the composite models are not affected by it.
 
\begin{figure*} 
\includegraphics[width=0.49\textwidth]{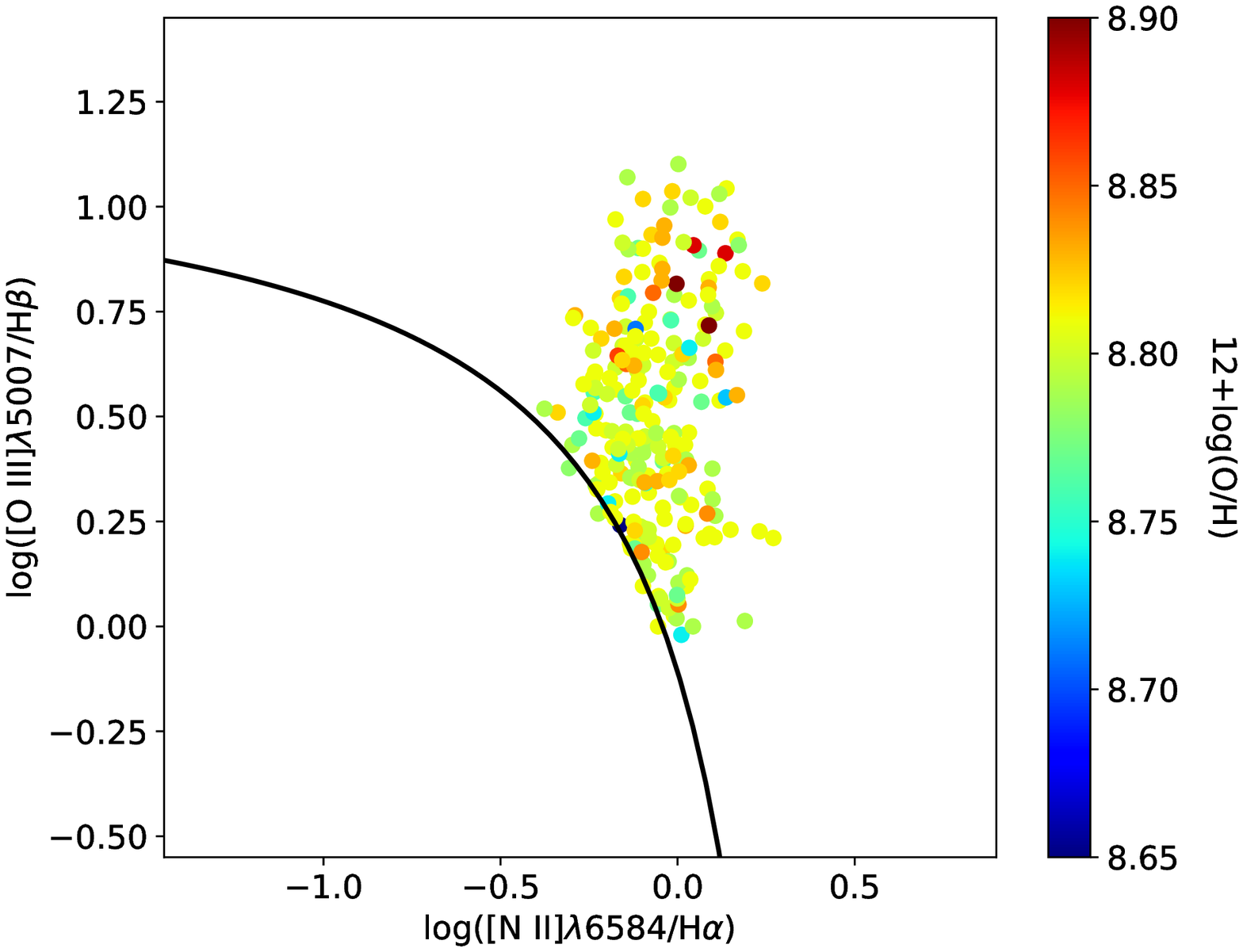}
\includegraphics[width=0.49\textwidth]{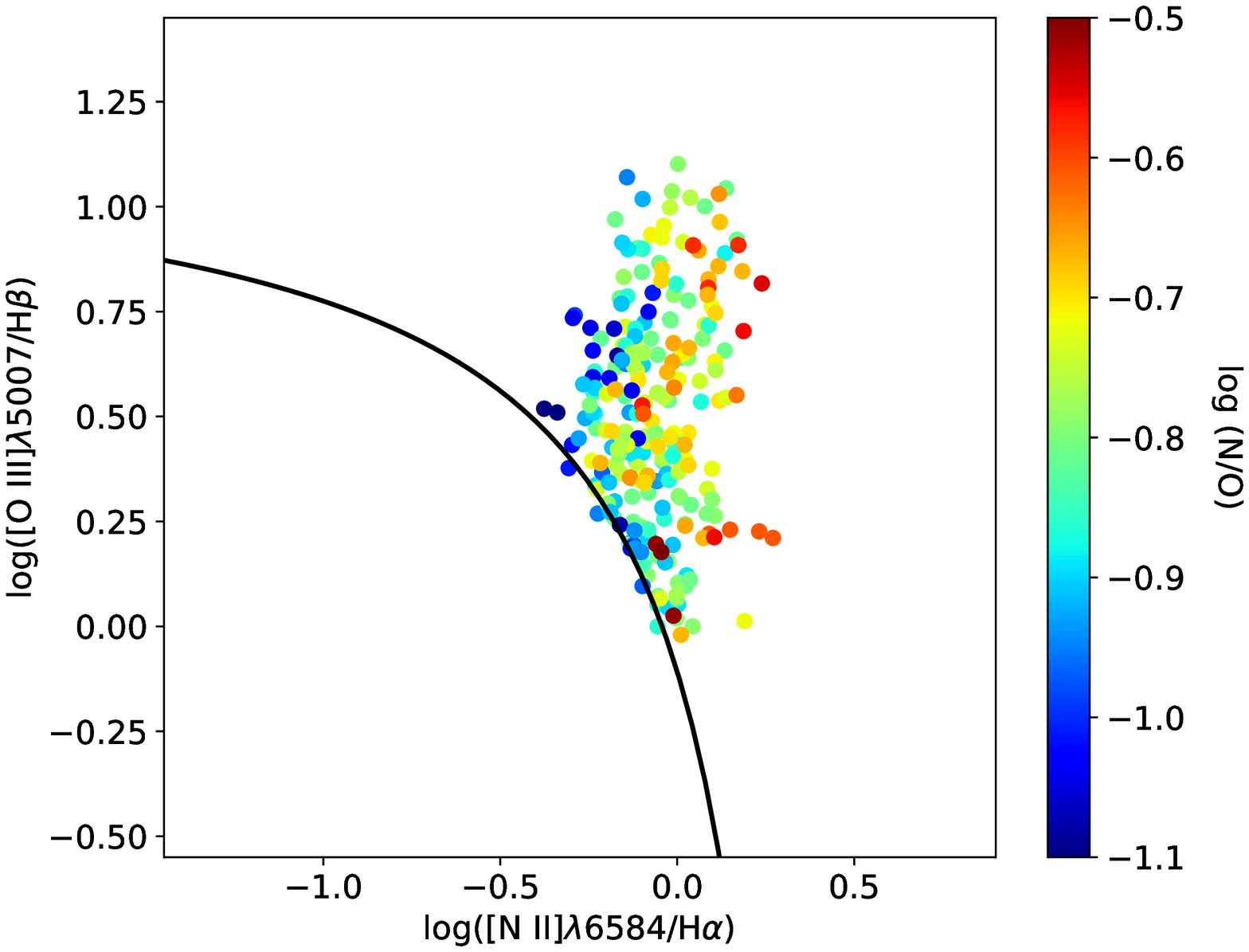}
\caption{Diagnostic diagram [\ion{O}{iii}]$\lambda$5007/H$\beta$
versus [\ion{N}{ii}]$\lambda$6584/H$\alpha$. The points represent objects of our sample
and the color bars show  the predicted composite model values of 
12+log(O/H) (left panel) and log(N/O) (right panel) abundance ratios.
Line represents the  criterion proposed by \citet{kewley01}  to separate
AGN-like and \ion{H}{ii}-like objects.}
\label{fig7}
\end{figure*}

In Fig.~\ref{fig7}, we verify the position of our sample objects in the [\ion{O}{iii}]$\lambda$5007/H$\beta$
versus [\ion{N}{ii}]$\lambda$6584/H$\alpha$ diagnostic diagram in terms of the abundance
ratios 12+log(O/H) (left panel). In disagreement with
pure photoionization model results (e.g. \citealt{feltre16}), it is not possible to distinguish
objects with different O/H values in this diagram. Our 
models predict  a narrow range (see above)  of O/H abundances and the  large majority of our sample
objects ($\sim99$ per cent) shows $\rm 12+\log(O/H)$ values higher than 8.7 dex. A narrow metallicity range for AGNs, with
12+log(O/H) varying  by $\sim 0.1$ dex as a function of host galaxy stellar mass over
the range  $10.1 \: \la \: \log(M/ \rm M_{\odot}) \: \la \: 11.3$, was also derived by
\citet{thomas19}, who used  the Bayesian parameter estimation code {\sc NebulaBayes}.
The derived narrow O/H range by the composite models for our sample  reflects  a stronger dependence of
 oxygen emission lines on shock parameters (e.g. \Vs) 
 rather than the O/H abundance. Moreover, there is a 
 degeneracy in the models which is also found in pure photoionization models.
In fact, \citet{davies14b} who considered the  Starburst-AGN mixing models, pointed out that
 metallicity and ionization parameters are degenerated quantities  because pure photoionization models adopting 
 different combinations
of these parameters can produce similar line ratios. In our case, models  adopting different \Vs, $n_{0}$
and metallicities can produce  similar emission-line intensities.
The degeneracy disappears when many lines from different ionization levels are observed for each element
in single spectra.

The  right panel of Fig.~\ref{fig7} shows the  BPT diagram for our sample in terms of different N/O abundance values, 
represented by the different colors of the points.
 There is a clear trend showing that objects with higher log(N/O) have also higher
[\ion{N}{ii}]$\lambda$6584/H$\alpha$  values. Although expected (see \citealt{ji20}),  this result is very interesting,
because the  interpretations of observation data of AGNs
based on shock models, in general,  consider a fixed value for the metallicity (or oxygen abundance).
Moreover, this result indicates that  our composite models are able to estimate N/O abundances by using
nitrogen lines, less sensitive to \Vs. Other diagnostic diagrams involving [\ion{O}{i}]$\lambda$6300/H$\alpha$
and [\ion{S}{ii}]$\lambda$6725/H$\alpha$  are not considered as these line ratios show a low dependency
with O/H and N/O abundance ratios.

\begin{figure*} 
\includegraphics[scale=0.36, angle=0]{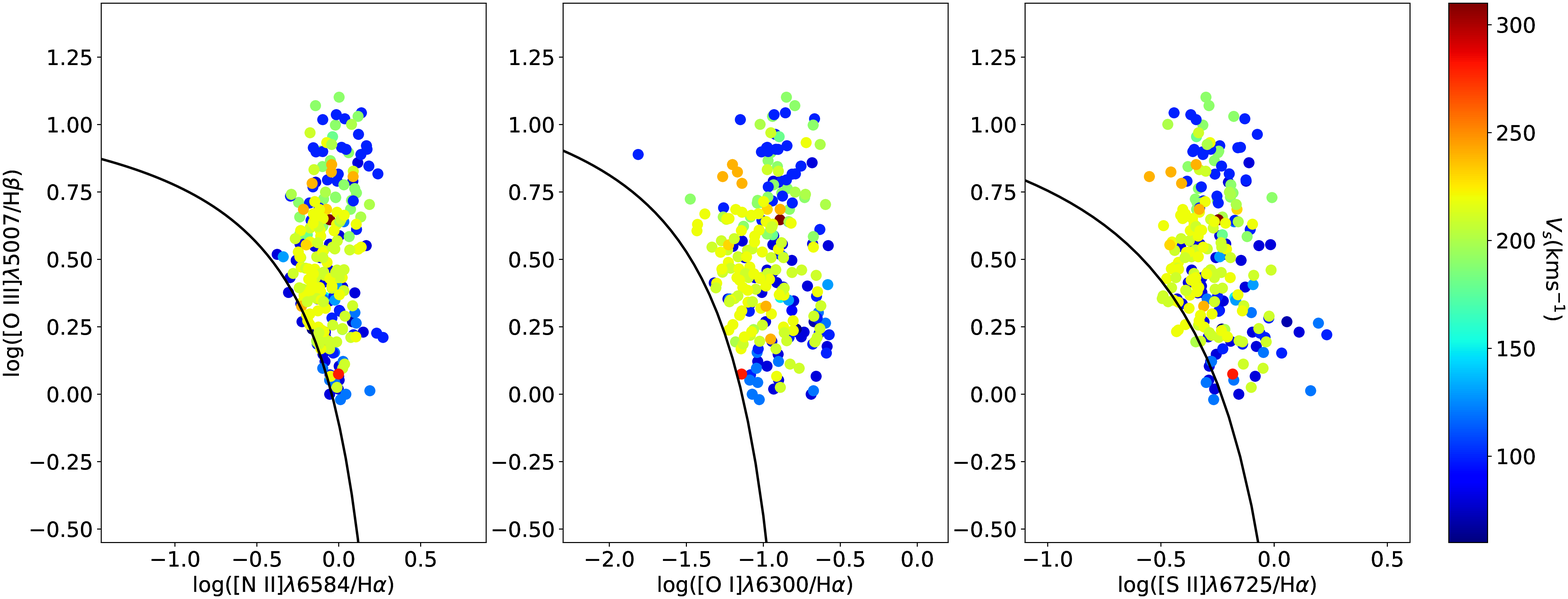}
\caption{Diagnostic diagrams [\ion{O}{iii}]$\lambda$5007/H$\beta$
versus [\ion{N}{ii}]$\lambda$6584/H$\alpha$, versus [\ion{O}{i}]$\lambda$6300/H$\alpha$
and versus [\ion{S}{ii}]$\lambda$6725/H$\alpha$. The points represent objects of our sample
(see Sect.~\ref{obs}) separated  according to the  
shock velocities (\Vs~ in units of $\rm km\: s^{-1}$) predicted by our composite models.
The lines represent the  criteria proposed by \citet{kewley01} and \citet{enrique13} to separate
AGN-like from \ion{H}{ii}-like objects.}
\label{fig8}
\end{figure*}

In Fig.~\ref{fig8}, three diagnostic diagrams, [\ion{O}{iii}]$\lambda$5007/H$\beta$
versus [\ion{N}{ii}]$\lambda$6584/H$\alpha$,  [\ion{O}{i}]$\lambda$6300/H$\alpha$
and  [\ion{S}{ii}]$\lambda$6725/H$\alpha$, the objects of our sample are separated
 according to the predicted composite model results for  the
shock velocity \Vs. There is no correlation between the position of the objects
and \Vs. The highest values of [\ion{O}{iii}]$\lambda$5007/H$\beta$ correspond
to \Vs\, in the range 100--200 $\rm km\: s^{-1}$. This is due, for models with $V_{\rm s} \: \ga \: 200 \: \rm km\: s^{-1}$, to the $\rm O^{3+}$
ionic abundance increase and, consequently, to the decrease of the lines emitted by $\rm O^{2+}$ ion, as already reported
above. One direct consequence of this result is that standard diagnostic diagrams, 
based on integrated spectra, can not be
used to distinguish the shock velocities in AGNs.

\begin{figure} 
\includegraphics[scale=0.6, angle=-90]{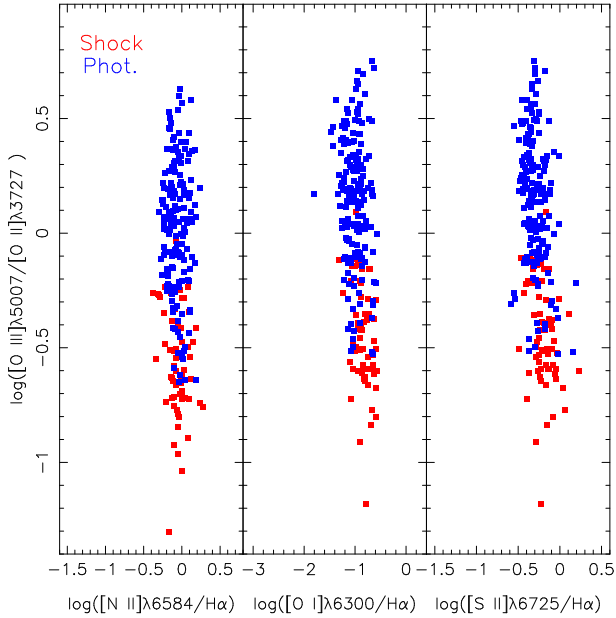}
\caption{Logarithm of the line ratios [\ion{O}{iii}]$\lambda$5007/[\ion{O}{ii}]$\lambda$3727
versus [\ion{N}{ii}]$\lambda$6584/H$\alpha$, [\ion{O}{i}]$\lambda$6300/H$\alpha$
and [\ion{S}{ii}]$\lambda$6725/H$\alpha$. Red and blue points represent objects of our sample
classified as shock ($F \rm \: < \: 10^{9} \: ph \: cm^{-2} \: s^{-1} \: eV^{-1}$) 
and photoionization dominated, respectively.}
\label{fig9}
\end{figure}

\begin{figure} 
\includegraphics[scale=0.6, angle=-90]{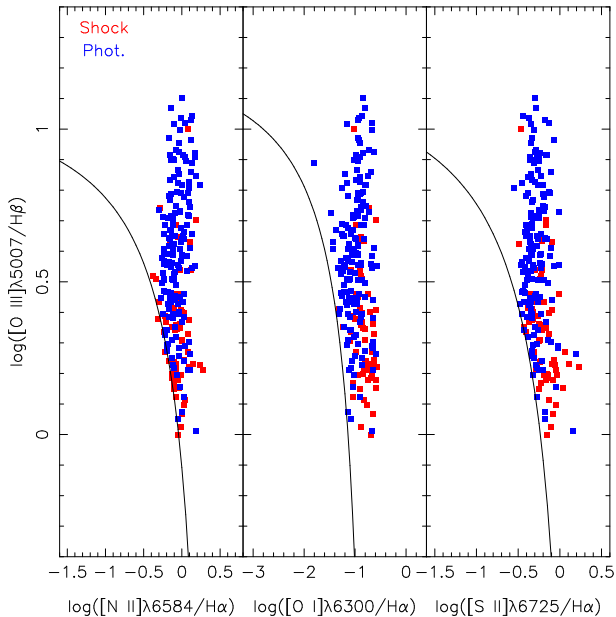}
\caption{As Fig.~\ref{fig9} but for the 
logarithm of the line ratios [\ion{O}{iii}]$\lambda$5007/H$\beta$  versus
 [\ion{N}{ii}]$\lambda$6584/H$\alpha$,  [\ion{O}{i}]$\lambda$6300/H$\alpha$
and  [\ion{S}{ii}]$\lambda$6725/H$\alpha$.}
\label{fig10}
\end{figure}

Another important issue is to investigate the positions in the BPT diagrams of the objects whose  dominant ionization 
 mechanism  is shock and to compare their position with
that of photoionized objects. As previously reported, AGNs with
$F \rm \: < \: 10^{9} \: ph \: cm^{-2} \: s^{-1} \: eV^{-1}$ are
 are considered as being shock-dominated, otherwise they are considered as being photoionization-dominated. The   
shock-dominated regime is characterised by  log([\ion{O}{iii}]$\lambda$5007/[\ion{O}{ii}]$\lambda$3727) $\la \:0$,
as pointed out by \citet{contini12}, who used composite models to reproduce
the continuum and optical lines   emitted from the extended narrow-line region (ENLR) of the Seyfert 2 galaxy
NGC\,7212. In Fig.~\ref{fig9}, we show the logarithm of the line ratios [\ion{O}{iii}]$\lambda$5007/[\ion{O}{ii}]$\lambda$3727  versus
 [\ion{N}{ii}]$\lambda$6584/H$\alpha$,  [\ion{O}{i}]$\lambda$6300/H$\alpha$
and  [\ion{S}{ii}]$\lambda$6725/H$\alpha$ for our sample, splitted in terms of  shock-dominated and 
photoionization-dominated. Although an overlap of the line ratios from
the different regimes can be seen, the highest and the lowest [\ion{O}{iii}]/[\ion{O}{ii}] are only
observed in photoionization and shock dominated objects, respectively. In Fig.~\ref{fig10},
the shock and photoionization dominated objects of our sample are plotted in the BPT
diagrams. Although there is some overlap,  photoionization-dominated
models present higher [\ion{O}{iii}]/H$\beta$ values while the shock-dominated models present lower values.

\subsection{Chemical abundances}

\begin{figure} 
\includegraphics[scale=0.6, angle=-90]{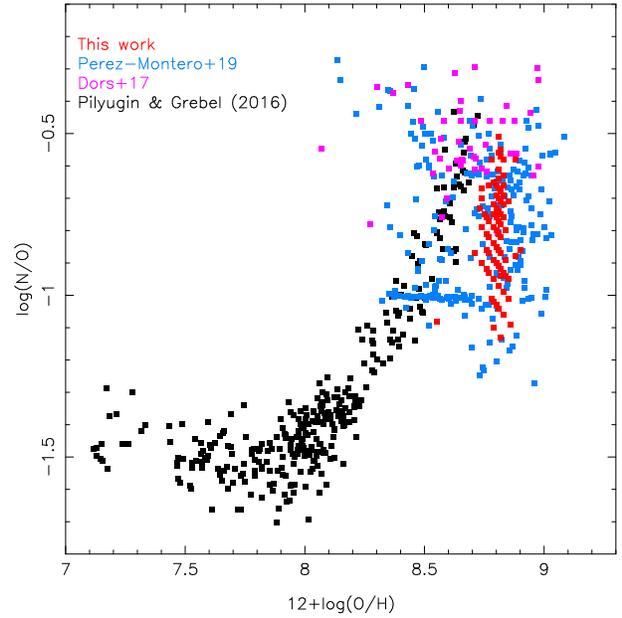}
\caption{log(N/O) versus 12+log(O/H) abundance ratio values. Red
points are values predicted by the {\sc composite} models (see Sect.~\ref{model}).
Blue points are  predicted values obtained by \citet{enrique19}
 using  the  {\sc \ion{H}{ii}-Chi-mistry} code  
for the same sample of objects considered in this work.
Pink points represent values  predicted by  detailed photoionization
models for a sample of 47 Seyfert 2   nuclei  ($z \: < 0.1$) by  \citet{dors17}.
Black points are estimations for \ion{H}{ii} regions   derived by   \citet{leonid16} adopting  the $C$ method \citep{leonid12}.}
\label{fig11}
\end{figure}

The heavy element abundances in  Seyfert~2   have been obtained in general only
for  oxygen and adopting  pure photoionization  models. In fact,
\citet{ferland83}, by using the first versions of the {\sc Cloudy} code \citep{ferland80},
showed that models employing a power-law ionizing continuum, metallicities
in the range $0.1 \: \lid \: (Z/{\rm Z_{\odot}}) \: \lid \: 1.0$ and ionization parameter
in the range $-4.0 \: \lid \: (\log U) \: \lid \: -2.0$
are able to describe  the sequence of the  optical emission-line ratios of Seyferts in BPT diagrams.
After this pioneering work many authors have invoked   pure photoionization models
to derive physical properties of AGN NLRs at low (e.g. \citealt{grazyna84, ferland86,
cruz91, thaisa98, groves06, feltre16, castro17, enrique19, sarita20}) 
and high redshifts (e.g. \citealt{nagao06a, matsuoka09, matsuoka18,  nakajima18, dors18, mignoli19, guo20}).
\citet{dors20a}  showed that the methods based on
pure photoionization models derive 12+log(O/H) values in the range from $\sim 7.2$ to $\sim 9.2$.
Among the methods considered by \citet{dors20a}, the results obtained through the {\sc \ion{H}{ii}-Chi-mistry} code 
\citep{enrique14}  can be used to compare abundances obtained  adopting pure photoionization with those based on composite
models. The {\sc \ion{H}{ii}-Chi-mistry} code was adapted for AGNs by \citet{enrique19} and  establishes a bayesian-like comparison 
between  predictions from a grid of photoionization models, built with the {\sc Cloudy} code \citep{ferland17},
and observational emission-line ratios. In view of  this, in Fig.~\ref{fig11}, the N/O versus O/H abundance values derived 
from  {\sc \ion{H}{ii}-Chi-mistry} code by \citet{enrique19} for the 244 objects of our sample are compared with those
derived by our composite models. In addition, in Fig.~\ref{fig11}, the values predicted by  individual photoionization
models for a different sample of 47 Seyfert~2 nuclei ($z \: < \: 0.1$) by  \citet{dors17},  obtained by the {\sc Cloudy} code,
and estimates for \ion{H}{ii} regions   derived by   \citet{leonid16}, who adopted the  $C$ method \citep{leonid12},
are shown. Despite the scattering,  the abundance results based on the  composite models
and those from the {\sc \ion{H}{ii}-Chi-mistry} code are located in the same region in  Fig.~\ref{fig11}.
However, the former predicts a narrower range of O/H values than those from the latter.
As can be seen in the  Fig.~\ref{fig11}, the points of the sample from \citet{dors17} ocuppy the 
region of highest metallicity,   which is problably due to the fact that  the  sample considered 
 in that work is  different  from  the one considered here, consisting mostly of strong AGNs, i.e., 
with very high [\ion{O}{iii}]$\lambda$5007/H$\beta$ ratios (see Fig.~1
of \citealt{dors20b}).

\begin{figure}
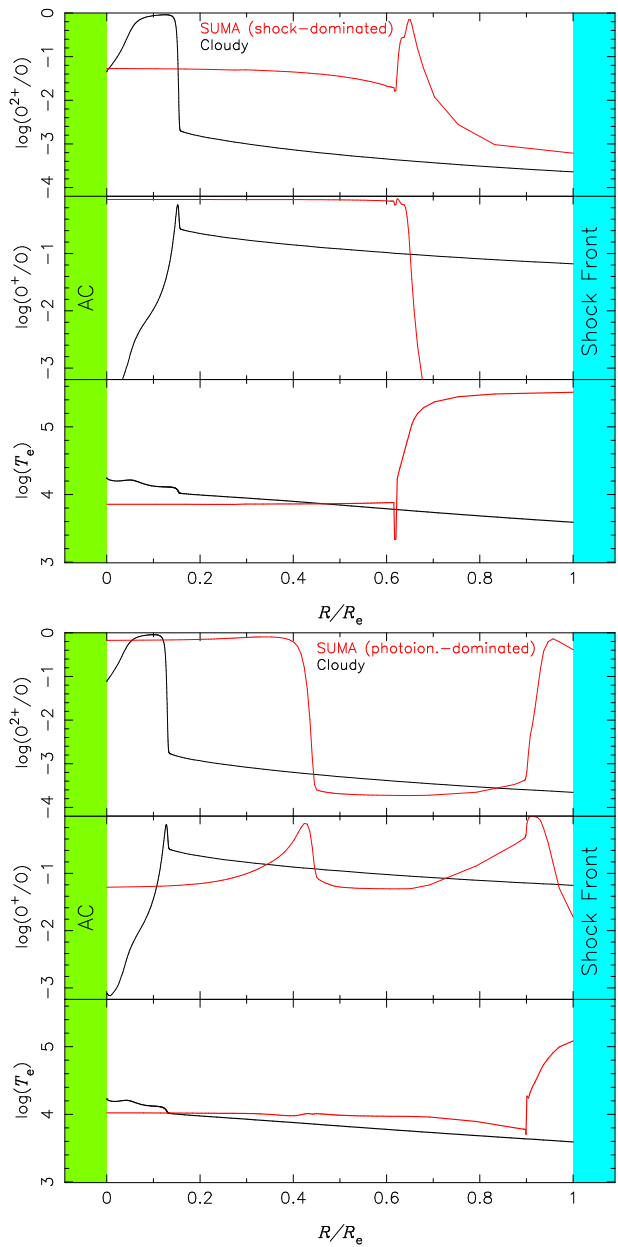
 
\includegraphics[scale=0.6, angle=-90]{ana2.eps}
\includegraphics[scale=0.6, angle=-90]{ana3.eps}
\caption{Profiles of electron temperature ($T_{\rm e}$) and  fractional abundances of the  oxygen ions
 O$^+$/O and O$^{++}$/O predicted by   {\sc SUMA} and {\sc Cloudy} codes. 
 The distance $R$ from the edge reached by radiation from active centre (AC) was normalised by the outermost 
radius $R_{\rm e}$ of each model.
The positions of the  edge illuminated by the  AC radiation  and of  the shock front (for {\sc SUMA} models)
 are indicated in each plot.  Top panel: Profiles for a 
 shock dominated model (m6). The  parameter set [12+log(O/H), \Vs\, ($\rm km\: s^{-1}$), $n_{0}$ ($\rm cm^{-3})$, log $F$]
 for the {\sc SUMA} model are (8.81, 150, 90, 8.41). The {\sc Cloudy} model was obtained from \citet{enrique19} 
 assuming the parameter set [12+log(O/H), $N_{\rm e}$ ($\rm cm^{-3})$, $\alpha_ox$, $\log U$] equal to 
 [8.82, 500, $-0.8$, $-2.0$]. Bottom panel: As the top panel but for a photoionization dominated model (m26) with
 {\sc SUMA} and {\sc Cloudy} parameters  (8.81, 100, 160, 9.95) and [8.77, 500, $-0.8$, $-2.1$], respectively.}
\label{fig12}
\end{figure}

The difference between the O/H abundance results obtained by {\sc SUMA} and those obtained by {\sc CLOUDY}, i.e.
by the {\sc \ion{H}{ii}-Chi-mistry} code,
is due to the presence of shocks in our models,  explained as follows. 
In  case of ejection, the clouds move outwards.  The shock front is on the outer edge of the clouds, 
while the photoionization flux from AC reaches the opposite (internal) edge.
Therefore  different temperature profile and ionization structure throughout the clouds are predicted by the 
{\sc SUMA} and {\sc Cloudy} codes and, consequently, different O/H abundances result
because the  oxygen lines are strongly dependent on the shock parameters.
To  illustrate this point, we select from our results two models; one
predicted to be  shock dominated (m6) and another one which is  photoionization dominated (m26).
The set of parameters [12+log(O/H), \Vs\, ($\rm km\: s^{-1}$), $n_{0}$ ($\rm cm^{-3})$, log$F$]
for m6 and m26 are (8.81, 150, 90, 8.41) and (8.81, 100, 160, 9.95), respectively (see Table~\ref{tab2}). 
The {\sc Cloudy} models were obtained by
\citet{enrique19} and the set of parameters [12+log(O/H), $N_{\rm e}$ ($\rm cm^{-3})$, $\alpha_{ox}$, $\log U$]
for m6 and m26 are [8.82, 500, $-0.8$, $-2.0$] and [8.77, 500, $-0.8$, $-2.1$], respectively.
In Fig.~\ref{fig12}, we illustrate the profiles of \Te ~ and of the fractional abundances of the  oxygen ions
 O$^+$/O and O$^{2+}$/O throughout the clouds of m6 and m26 
in order to understand the  [\ion{O}{ii}] and [\ion{O}{iii}] line intensity results.
The distance $R$ from the edge illuminated by the AC radiation  within the cloud  was normalised by the outermost 
radius $R_{\rm e}$ of each model.
In order to  clarify the figure interpretation, the  radiation illuminated  and the  shock front (for the {\sc SUMA} 
models) 
positions are indicated in Fig.~\ref{fig12}. For  both {\sc SUMA} results, one can see that $T_{\rm e}$
reaches a high value ($T_{\rm e}\sim \: 10^{5}$ K) downstream  near the the shock front.  However, for 
the photoionization dominated model,  $T_{\rm e}$  reaches about the same value predicted by the {\sc Cloudy} model
at $(R/R_{\rm e}) \: \la \: 0.9$. Otherwise, for the shock dominated model,
high $T_{\rm e}$ values extend to about half  of the radius. Concerning the O$^+$/O fractional abundance,
both photoionization and  shock dominated models produce different profiles  to those calculated  by the {\sc Cloudy},
with the shock dominated model producing  very small fractional abundance for $(R/R_{\rm e}) \: \ga \: 0.6$.
Finally, O$^{2+}$/O structures predicted by both {\sc SUMA} models indicate a higher level
of ionization along the radius as compared to those calculated by the {\sc Cloudy} models, 
being  more  pronounced for the shock dominated model.
Therefore, any physical property derived through [\ion{O}{ii}], [\ion{O}{iii}] and
high ionization lines (e.g. [\ion{Fe}{vii}]$\lambda$6087, [\ion{Fe}{x}]$\lambda$6375, etc.)
can be very different when composite  or pure photoionization models are adopted.

\subsection{Abundance calibrations}

\citet{thaisa98} proposed the first calibration between the metallicity (in terms of O/H)
and  narrow optical emission line ratios of AGNs. After this pionering work, \citet{castro17}
and \citet{sarita20} proposed semi-empirical calibrations based on the line
ratios $N2O2$=log(\ion{N}{ii}$\lambda$6584/\ion{O}{ii}$\lambda$3727) and 
$N2$=log(\ion{N}{ii}$\lambda$6584/H$\alpha$), respectively. In particular, \citet{castro17}
pointed out the importance  of considering emission lines   emitted by ions with similar 
ionization potential, in order to minimize the shock effects on  calibrations.
In fact, as shown in Fig.~\ref{fig12}, the inclusion of shocks  in the 
gas ionization rate, in addition to photoionization from AGN, produces  different 
temperature and fractional abundance profiles, mainly for  shock dominated objects.
In this sense, the $N2O2$ index  has an advantage relative to other line ratios because the involved ions, $\rm N^{+}$
and $\rm O^{+}$, have  close ionization potentials, i.e. 29.60 eV and 35.12 eV, respectively.

\begin{figure} 
\includegraphics[scale=0.6, angle=-90]{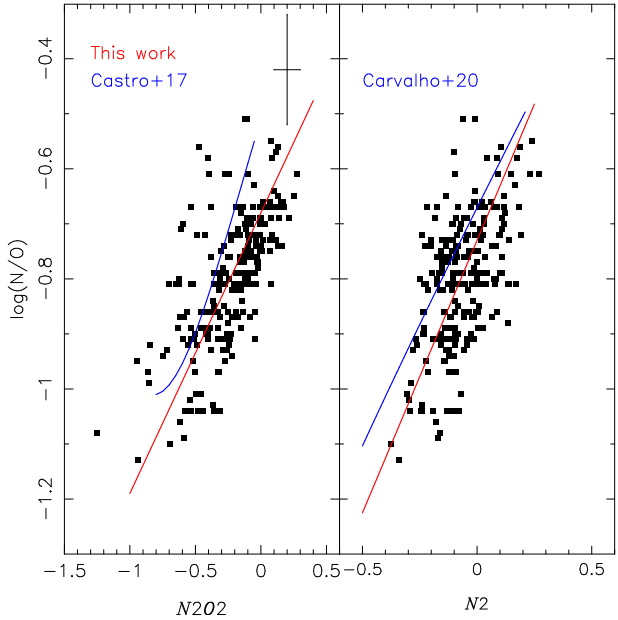}
\caption{Left panel: Logarithm of the abundance ratio N/O versus $N2O2$=log(\ion{N}{ii}$\lambda$6584/\ion{O}{ii}$\lambda$3727).
Points represent the log(N/O) predicted by the composite models (see Sect.~\ref{model}) and the corresponding
observational $N2O2$ of the sample  objects  (see Sect.~\ref{obs}). Red line represents the linear fit to the 
points given by Eq.~\ref{eqcsh}. Blue line represents the (N/O)-$N2O2$ derived by using the
calibration proposed by \citet{castro17} (Eq.~\ref{cast17}, Eqs.\ref{oxz} and \ref{nod17}).
Error bars represent the observational error of  $N2O2$ and the uncertainty in the estimation of log(N/O) (e.g. \citealt{denicolo02}).
Rigth panel: Same as left panel but for log(N/O) versus $N2$. Red line represents the linear fitting to the 
points given by Eq.~\ref{nod99}. Blue line represents the (N/O)-$N2O2$ derived by using the
calibration proposed by \citet{sarita20} (Eq.~\ref{eqsart}) and  Eqs.~\ref{oxz} and \ref{nod17}. }
\label{fig13}
\end{figure}

All  calibrations between $Z$ and strong emission-lines for AGNs available in the literature 
(for a review, see \citealt{dors20a}) are based on pure photoionization models
and  it is thus worth to obtain  new calibrations considering the contribution from
shocks to the NLR.
Therefore, in Fig.~\ref{fig13}, the predicted N/O abundance ratio versus the
observational $N2O2$ value for each object of our sample is shown.  A
clear correlation  can be seen between the points (with the Pearson correlation coefficient equal to 0.65) and it is represented by 
\begin{equation}
\label{eqcsh}
\log({\rm N/O})= (0.51\pm0.02) \times \: N2O2 + (-0.68\pm0.01).   
\end{equation}
Similarly, a (N/O)-$N2O2$ calibration was proposed by \citet{enrique09} for 
\ion{H}{ii} regions. 

 Considering that this is the first
metallicity calibration derived  using composite models,  
 a  comparison with  calibrations proposed
by other authors should be presented. The unique calibration which uses the
$N2O2$ index as metallicity indicator for AGNs  has been 
 proposed by  \citet{castro17}, 
\begin{equation}
\label{cast17}
(Z/{\rm Z_{\odot}})=1.08 \times N2O2^{2}+ 1.78 \times N2O2 +1.24.
\end{equation}
 
Nevertheless, this calibration considers  the metallicity and our calibration the relative
abundance N/O ($\sim Z$, \citealt{vila93, henry00}). Therefore, to consistently compare
both calibrations the following procedure is carry out. 
First, we  convert the 
$(Z/{\rm Z_{\odot}})$ values in O/H abundances assuming
\begin{equation}
\label{oxz}
{\rm 12+log(O/H)}=12+\log[(Z/{\rm Z_{\odot}})\rm \: \times \: 10^{log(O/H)_{\odot}}],
\end{equation} 
where  $\rm log(O/H)_{\odot}=-3.31$ \citep{allendeprieto01}. Then, we
take into account the (N/O)-(O/H) relation derived  by \citet{dors17},
obtained by using abundance estimates  of \ion{H}{ii} regions
and Seyfert~2, given by
\begin{equation}
\label{nod17}
\rm \log(N/O)=1.29 \: \times 12+\log(O/H) - 11.84.
\end{equation}
 The resulting (N/O)-($N2O2$)  \citet{castro17} calibration
is compared  to our calibration (Eq.~\ref{eqcsh})
in Fig.~\ref{fig13}, left panel, where a good agreement can be  observed, taking
into account the   observational error of $N2O2$  ($\sim$0.1 dex)
and the uncertainty of $\sim$0.2 dex in estimations based
on strong-line methods (e.g. \citealt{denicolo02}). 

 Taking into account that the $N2O2$ parameter could be affected 
by   reddening correction, we also explore the relation with $N2$
parameter that is not affected by reddening correction. In Fig.~\ref{fig13}, 
rigth panel, we present the linear calibration between  $\rm \log(N/O)$ and $N2$, obtaining  
 \begin{equation}
\label{nod99}
\log({\rm N/O})= (0.99\pm0.05) \times \: N2 + (-0.73\pm0.01),   
\end{equation}
 with the Pearson correlation coefficient equal to 0.55.
\citet{sarita20} proposed a semi-empirical
calibration, based on the {\sc Cloudy} models, between $Z$ and $N2$ given by
\begin{equation}
\label{eqsart}
(Z/Z_{\odot})=4.01^{N2}-0.07.
\end{equation} 
We adopt the same procedure previously described  to convert
the metallicity values derived from  the calibration by
\citet{sarita20} into N/O and obtain a calibration
between this abundance ratio and $N2$. 
The resulting  calibration is compared with the one obtained
through the composite models (Eq.~\ref{nod99}) in Fig.~\ref{fig13}, rigth panel.
As in the case of the (N/O)-($N2O2$), both calibrations 
are in agreement each other taking into accout the uncertainties 
in estimates based on strong-line methods. 
Since the dispersion present in both panels are very 
similar, we can conclude that the $N2O2$ relation proposed in this 
work is not strongly affected by reddening correction.

\section{Conclusion remarks} 
\label{conc}

In this  work, by using the {\sc SUMA} code, we calculate  the photoionization and shock parameters suitable
to reproduce  the optical ($3000 \: < \: \lambda$(\AA) $< \: 7000$)  emission lines emitted  from the NLRs of 
244 Seyfert~2 nuclei  in the local universe ($z \: \la \: 0.4$). 
The observation data   
were taken from  the Sloan Digital Sky Survey (SDSS).
Based on the results of the detailed modelling, we found that  Seyfert~2 present
shocks in their NLRs with velocities in the range   $\sim 50$ to $\sim 300$ $\rm km \: s^{-1}$ and an
average value  $\sim 170$ $\rm km \: s^{-1}$. A narrower range of metallicities 
($0.6 \: \la \:  (Z/Z_{\odot}) \: \la \: 1.6$) 
than that  estimated from  pure photoionization models is  derived for the sample.
The standard diagnostic diagrams [\ion{O}{iii}]$\lambda$5007/H$\beta$
versus [\ion{N}{ii}]$\lambda$6584/H$\alpha$,  [\ion{O}{i}]$\lambda$6300/H$\alpha$
and  [\ion{S}{ii}]$\lambda$6725/H$\alpha$, based on integrated spectra,  can  be used to discriminate between shock and photoionization
dominated objects.  However, our results indicate that shock velocity in
AGNs can not be estimated by these  standard optical line ratio diagrams.
Also, our results show that the 
 temperature structure and O$^+$/O and O$^{2+}$/O fractional abundance profiles along the radius of the emitting nebula
 are highly modified  by the shock presence.
These results  suggest that  a combination of 
lines emitted by ions with similar  ionization potential, in order to minimize the shock effects, 
are preferable to other metallicity indicators.
Finally, from our model results it was possible to derive   calibrations between the  N/O abundance ratio
and the $N2O2$=log(\ion{N}{ii}$\lambda$6584/\ion{O}{ii}$\lambda$3727) and $N2$=log(\ion{N}{ii}$\lambda$6584/H$\alpha$)
indexes. These calibrations are in
agreement  with  those derived from  pure photoionization models.

\section*{Acknowledgments}
We are grateful  to the referee for his/her dedicated work in
reviewing our paper.
OLD and ACK are grateful to  Funda\c c\~ao de Amparo \`a
Pesquisa do Estado de S\~ao Paulo (FAPESP) and Conselho Nacional
de Desenvolvimento Cient\'{\i}fico e Tecnol\'ogico (CNPq).
R.A.R. thanks partial financial support from Funda\c c\~ao de Amparo \`a Pesquisa do 
Estado do Rio Grande do Sul (17/2551-0001144-9 and 16/2551-0000251-7) and Conselho Nacional de 
Desenvolvimento Cient\'{\i}fico e Tecnol\'ogico (302280/2019-7).
MVC and GFH are grateful to CONICET.
 
\section{DATA AVAILABILITY}
The data underlying this article will be shared on reasonable request
to the corresponding author.


\label{lastpage}


\end{document}